\documentclass[]{imsart}

\pdfoutput=1

\usepackage{graphicx}
\usepackage{caption}
\usepackage{subcaption}

\usepackage{array}

\usepackage{amssymb}
\usepackage{amsfonts}
\usepackage{multirow}
\usepackage{bigstrut}

\RequirePackage[OT1]{fontenc}
\RequirePackage{amsthm,amsmath}

\RequirePackage{natbib}
\RequirePackage[colorlinks,citecolor=blue,urlcolor=blue]{hyperref}

\startlocaldefs

\newcommand{\matr}[1]{\mathbf{#1}}
\newcommand{\vect}[1]{\mathbf{#1}}
\newcommand{\el}{ {\cal{L}}}
\newcommand{\E}{ {\rm{E}}}
\newcommand{\I}{ {\rm{I}}}

\newcommand{\Covfj} {{\mathbf{A}(\lambda_j)}}

\newcommand{\pijc} {{p_{ij}^{(c)}}}
\newcommand{\minitab}[2][l]{\begin{tabular}{#1}#2\end{tabular}}

\newcommand{\thickhline}{%
    \noalign {\ifnum 0=`}\fi \hrule height 1pt
    \futurelet \reserved@a \@xhline
}

\endlocaldefs

\begin{document}

\begin{frontmatter}

\title{Switching Nonparametric Regression Models and the Motorcycle Data revisited}
\runtitle{Switching Nonparametric Regression Models}


\author{\fnms{Camila} \snm{P. E. de Souza}\thanksref{t1}\ead[label=e1]{camila.souza@stat.ubc.ca}}

\and
\author{\fnms{Nancy} \snm{E. Heckman}\thanksref{t1}\ead[label=e2]{nancy@stat.ubc.ca}}
\address{Department of Statistics - The University of British Columbia \\
3182 Earth Sciences Building, 2207 Main Mall\\
Vancouver, BC Canada V6T 1Z4\\
\printead{e1}\\
\printead*{e2}
}
\affiliation{University of British Columbia}
\thankstext{t1}{Research supported by the National Science and Engineering Research Council of Canada.}

\runauthor{de Souza and Heckman}

\begin{abstract}
We propose a methodology to analyze data arising from a curve that, over its domain, switches among $J$ states. We consider a sequence of response variables, where each response $y$ depends on a covariate $x$ according to an unobserved state $z$. The states form a stochastic process and their possible values are $j=1,\ldots,J$. If $z$ equals $j$ the expected response of $y$ is one of $J$ unknown smooth functions evaluated at $x$. We call this model a switching nonparametric regression model. We develop an EM algorithm to estimate the parameters of the latent state process and the functions corresponding to the $J$ states. We also obtain standard errors for the parameter estimates of the state process. We conduct simulation studies to analyze the frequentist properties of our estimates. We also apply the proposed methodology to the well-known motorcycle data set treating the data as coming from more than one simulated accident run with unobserved run labels.
\end{abstract}




\end{frontmatter}


\section{Introduction}

In this paper we propose a methodology to analyze data arising from a curve that, over its domain, switches among $J$ states. The state at any particular point is determined by a latent process. The state also determines a function. We are interested in the functions corresponding to each of the states and the parameters of the latent state process.

Suppose we have a sequence of response variables, $y_1,\ldots,y_n$, where $y_i$ depends on a covariate $x_i$ (usually time) according to an unobserved state $z_i$, also called a hidden or latent state. The possible values of the states are $j=1,\ldots,J$. If $z_i=j$ the expected response of $y_i$ is $f_j(x_i)$. We call this model a switching nonparametric regression model. In a Bayesian switching nonparametric regression model the uncertainty about the $f_j$'s is formulated by modeling the $f_j$'s as realizations of stochastic processes. In a frequentist switching nonparametric regression model the $f_j$'s are merely assumed to be smooth.

The objective of the proposed methodology is to estimate the $f_j$'s as well as to estimate the regression error variance and the parameters governing the distribution of the state process. We also obtain standard errors for the proposed parameter estimators of the state process. We consider two types of hidden states, those that are independent and identically distributed and those that follow a Markov structure. The Markov structure would usually require $x$ to be time.

\begin{figure}
\begin{center}
\includegraphics[width=11cm]{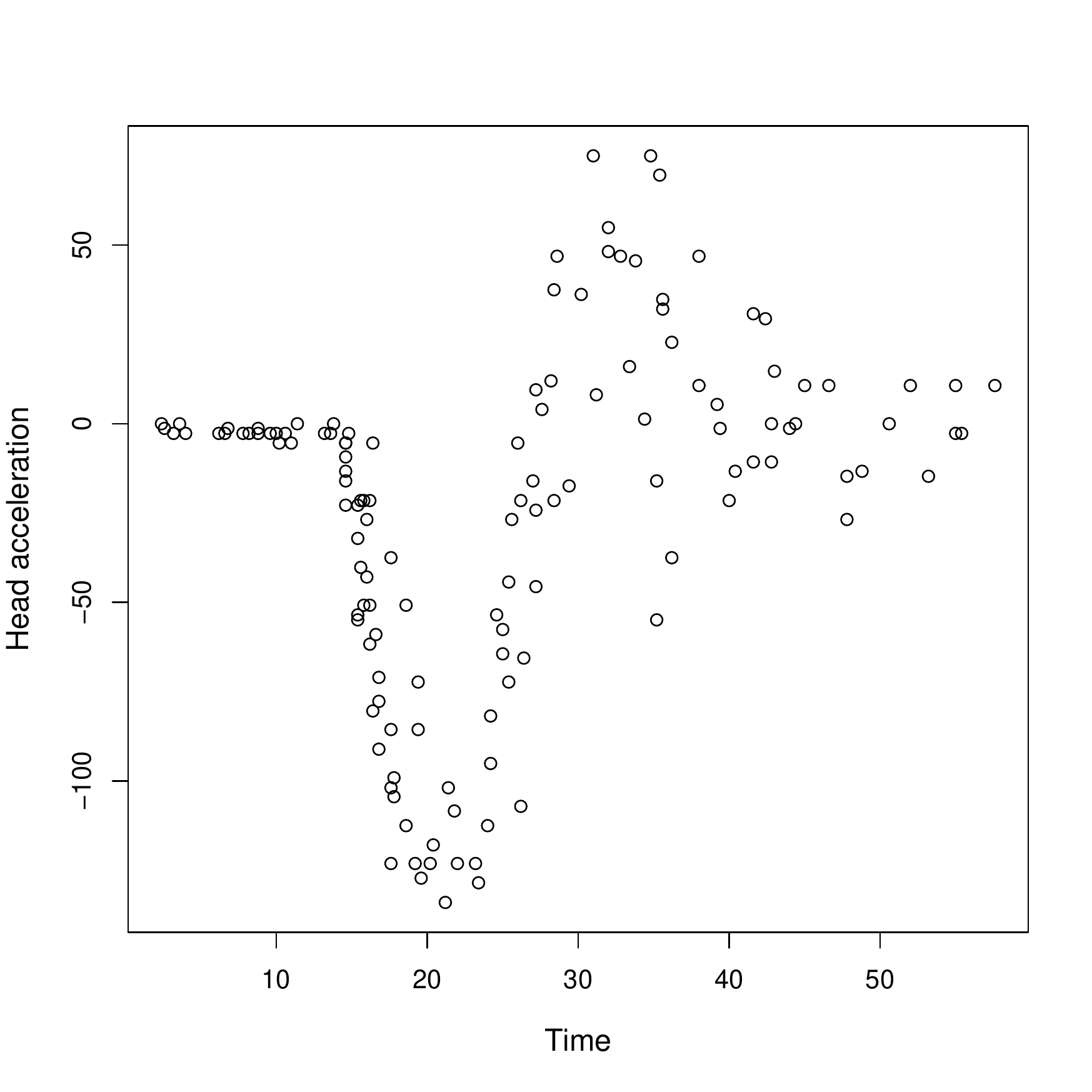}
\caption{Motorcycle data. Head acceleration in $g$ (one $g \approx 9.8m /s^2$) versus the time in milliseconds after impact.}
\label{motorcycle}
\end{center}
\end{figure}

As an application we consider the well-known motorcycle data set. The data consist of 133 measurements of head acceleration taken through time in a simulated motorcycle accident. See Figure \ref{motorcycle}. Analyses appearing in the literature (e.g., \citealp{silverman1985some}, \citealp{hardle1995fast}, \citealp{rasmussen2002}, \citealp{gijbels2004} and \citealp{wood2011fast}) treat the data as coming from one simulated accident. However, close examination of the data suggests the measurements are from $J >1$ accidents. In the discussion of Silverman (1985) Prof. A. C. Atkinson wrote ``inspection of [the Figure], particularly in the region 30-40 ms, suggests that the data may not be a single time series but are rather the superposition of, perhaps, three series". Professor Silverman had no specific information on this point but replied ``Professor Atkinson is right about the motorcycle data in that they are the superposition of measurements made by several instruments but the data I have presented are precisely in the form they were made available to me". The data structure will most likely remain unclear as the original report \citep{schmidt1981} seems to be no longer available.

We apply the proposed methodology to the motorcycle data treating the data as coming from $J$ functions, one for each simulated accident, with hidden (unobserved) function labels. We choose $J$ using an ad hoc Akaike information criterion (AIC). To our knowledge this is the first time that the motorcycle data is analyzed taking into account that they describe more than one simulated accident.

The paper is organized as follows. In Section \ref{overview} we present an overview of the proposed methodology. A literature review on the topic is presented in Section \ref{sec:background}. In Section \ref{EMsection} we describe the solution for the estimation problem. The standard errors for the parameter estimators of the state process are calculated in Section \ref{sec:standarderrors}. In Section \ref{sec:implem} we present the results of simulation studies. An application of the proposed methodology to the motorcycle data is shown in Section \ref{sec:application}. Some discussion is presented in Section \ref{sec:discussion}.

\section{Overview of the proposed methodology}\label{overview}
Suppose we have observed data $x_i,y_i$, and hidden states $z_i$, $i=1,\ldots,n$.  The states form a stochastic process and their possible values are $j=1,\ldots,J$. If $z_i = j$, then $y_i$'s distribution depends on a function $f_j$. Specifically, we assume that, given $z_i$, $y_i =  f_{z_i}(x_i) + \sigma_{z_i} \epsilon_{i}$, for $\epsilon_1,\ldots,\epsilon_n$ independent with mean of zero and variance of one. Therefore, given the $f_j$'s and the $z_i$'s, the $y_i$'s are independent with the mean of $y_i$ equal to $f_{z_i}(x_i)$ and the variance equal to $\sigma^2_{z_i}$.

Let $\vect{y}=(y_1,\ldots,y_n)^T$, $\vect{z}=(z_1,\ldots,z_n)^T$  and $f_j(\vect{x}) = (f_j(x_1),\ldots, f_j(x_n))^T$. Define $\gamma$ to be the set of parameters defining the distribution of $\vect{y}$ given $\vect{z}$ and the $f_j(\vect{x})$'s. So, for instance, if the $\epsilon_{i}$'s are normally distributed, then $\gamma=\{f_j(\vect{x})\; \mbox{and} \;\sigma_j^2 \; \mbox{for} \; j=1,\ldots,J\}$. Let $\alpha$ be the vector of parameters that determine the joint distribution of $\vect{z}$. If the $z_i$'s are independent and identically distributed, then the parameter vector $\alpha$ is of length $J$ with $j$th component equal to $P(z_i=j)$. If the $z_i$'s follow a Markov structure, the parameter vector $\alpha$ consists of initial and transition probabilities.

Our goal is to estimate $\theta \equiv \{\alpha,\gamma\}$, along with standard errors or some measure of accuracy for the parameters in $\alpha$.

To obtain the parameter estimates two approaches are considered:
\begin{itemize}
\item a frequentist approach called penalized log-likelihood estimation;
\item a Bayesian approach where the posterior density is maximized.
\end{itemize}

\noindent These two approaches are computationally similar and consist of using the Expectation-Maximization (EM) algorithm \citep{mclachlan} to maximize the following criterion
\begin{equation}
\mbox{log-likelihood of the data} + P(f_1,\ldots,f_J,\lambda_1,\ldots,\lambda_J),
\label{criterion1}
\end{equation}

\noindent where the exact form of $P(f_1,\ldots,f_J,\lambda_1,\ldots,\lambda_J)$ depends on the considered approach.

In the frequentist approach we assume that $f_j''$ exists almost everywhere and $\int [f_j''(x) \big ]^2dx < \infty$ for $j=1,\ldots,J$ and maximize (\ref{criterion1}) with
$$P(f_1,\ldots,f_J,\lambda_1,\ldots,\lambda_J) = - \sum_{j=1}^J \lambda_j \int [f_j''(x)]^2dx.$$

\noindent  The $\lambda_j$'s are the so called smoothing parameters as they measure the relative importance of fit to the data, as measured by the log-likelihood of the data, and smoothness of the $f_j$'s, as quantified by the penalties $\int [f_j''(x)]^2dx$.

The Bayesian approach requires a prior distribution for the parameters in $\theta$. For the $f_j$'s, we consider a Gaussian process regression approach \citep{rasmussen2006} with $f_1,\ldots, f_J$ independent and $f_j$ a Gaussian process with mean function $\mu_j(x) \equiv 0$ and covariance function, $K$, that depends on a vector of parameters $\lambda_j$. We place a non-informative prior on the other parameters. Therefore, we can write
\begin{eqnarray*}
P(f_1,\ldots,f_J,\lambda_1,\ldots,\lambda_J) &=& - ~  \frac{J}{2} \log (2 \pi)  -\frac{1}{2} \sum_{j=1}^J  \log | \Covfj|  \\
&& - ~ \frac{1}{2} \sum_{j=1}^J  f_j(\vect{x})^T ~ \Covfj^{-1} ~f_j(\vect{x}),
\end{eqnarray*}

\noindent where $\Covfj$ is an $n \times n$ matrix with entries given by
$$[A(\lambda_j)]_{lm}=K(x_l,x_m;\lambda_j),$$

\noindent with $K$ known. In our applications and simulations we assume the following covariance function
\begin{equation}
K(x,t; U_j, s_j^2) = {\rm{cov}}(f_j(x),f_j(t)) = U_j \exp\left[ - \frac{ (x-t)^2 }{2 s_j^2} \right],
\label{cov}
\end{equation}

\noindent with parameter vector $\lambda_j = (U_j, s_j)'$. These parameters control the amount of smoothness of each curve $f_j$ and, as in the penalized log-likelihood approach, are called smoothing parameters.

We can extend our approach to consider $\mu_j$ not equal to the zero function.  For instance, we might take $\mu_j(x)$ as a linear combination of known basis functions. In this case, in addition to $\lambda_j$, we would have another parameter vector for $f_j$, namely the vector containing the basis functions' coefficients.

In both approaches we choose the values of the $\lambda_j$'s automatically by cross-validation.

\section{Background}\label{sec:background}

Similar models have appeared in the machine learning literature, where they are called mixture of Gaussian processes models.  See \citet{tresp}, \citet{rasmussen2002} and \citet{ou2008}.
The focus of these papers  is on analyzing data from an on-line process, with the goal being prediction of a single process.  The process is modeled as a mixture of  realizations of Gaussian processes and,  just as in our model,
 the mixture depends on hidden states.
\citeauthor{rasmussen2002} and \citeauthor{ou2008} take a hierarchical Bayesian approach, not only placing a Gaussian process prior on the
$f_j$'s, but
also placing a prior distribution on all parameters, including those that govern the Gaussian process.    \citeauthor{tresp} does not place a prior distribution on the parameters of the Gaussian process, but he does use Gaussian processes to model not only the functions themselves but also the latent process and the
regression error variance.
 To analyze data,  \citeauthor{tresp} uses the Expectation-Maximization (EM) algorithm to maximize the posterior density of the
 Gaussian processes given the data.  \citeauthor{rasmussen2002} and \citeauthor{ou2008}
use a Monte Carlo method to estimate the posterior distribution of the unknown functions and parameters.

A main difference between our model and the model in these three papers is in the distribution of the $z_i$'s.  All three papers begin by assuming that the $z_i$'s are independent, conditional on the parameters governing the latent process. In \citeauthor{rasmussen2002} and in \citeauthor{ou2008}, the process parameters are $p_j=P(z_i = j)$, $j\geq 1$, which are modeled using a Dirichlet distribution. \citeauthor{rasmussen2002} use a limit of a Dirichlet distribution in order to model an infinite number of possible hidden states, to avoid choosing the number of states. \citeauthor{ou2008} use a finite number of states to avoid computational complexity. Both papers use  an ad hoc modification of the distribution of the $z_i$'s  to allow $z_i$ to depend on $x_i$ in a smooth way;  see \citeauthor{rasmussen2002}'s  equation (5) and \citeauthor{ou2008}'s equation (13).   However, as remarked by \citeauthor{rasmussen2002} in the discussion, the properties of the resulting joint distribution of the $z_i$'s are not completely  known. \citeauthor{tresp}'s distributional assumptions are more straightforward:  he assumes that the distribution of $z_i$ depends on $x_i$ according to a logit model governed by a Gaussian process. None of these papers consider Markov $z_i$'s.

While the three papers contain methodology that can, in principle, lead to estimation of the $f_j$'s and the latent variable process parameters, the papers  focus  on estimation of just one function - the mixture.  Thus the resulting methodology is a form of variable bandwidth smoothing (see, for instance, Fan and Gijbels, 1992). In contrast, our goal is estimation of the  individual processes that make up the mixture and estimation of the parameters governing the hidden state process, along with standard errors. We see the distinction between the goals  by considering the analysis of the motorcycle data:  \citeauthor{rasmussen2002}  present just one function to summarize the data.  We present $J>1$ functions, one for each of the $J$ simulated accidents, and we   estimate the expected proportion of data points  from each function, and provide standard errors. We also conduct extensive simulation studies of the frequentist properties of the estimates.

A closely related model is the Gaussian mixture model, used in density estimation. In the Gaussian mixture model, we assume that the $i$th data point comes from one of a finite set of Gaussian distributions, determined by the value of a latent variable $z_i$.
In his work in this area, \citet{bilmes1998} considered two models for the latent variables: in the first model the latent variables are
independent and identically distributed and in the second they follow a Markov structure. The later corresponds to a classic hidden Markov model, which is the same as our approach for Markov $z$'s if the $f_j$'s are constant. \citeauthor{bilmes1998}  provides a very readable description of how he applies the EM algorithm  to estimate the parameters for the Gaussian distributions as well as for the distribution of the $z_i$'s.

These  models are not to be confused with the work of \citet{shi2005}~on Gaussian process mixtures for regression and \citet{chiou2012} on functional mixture prediction. These authors analyze data from
$M$ independent curves where the entire $i$th curve is a realization of one of $J$  Gaussian processes, determined by the value of the latent variable $z_i$. In contrast, like \citeauthor{tresp}, \citeauthor{rasmussen2002} and \citeauthor{ou2008}, we consider $M=1$ observed curve, which switches among $J$ Gaussian processes.

We see that the literature contains many similar but distinct models with names containing the words {\em{Gaussian}} and {\em{mixture}}.   For this reason, we  prefer to call our models and those of \citet{tresp}, \citet{rasmussen2002} and \citet{ou2008}  {\em{switching nonparametric regression models}} as we feel this is more descriptive.
To our knowledge, no one has considered a Markov structure for the latent variables to estimate multiple functions, nor has anyone used the non-Bayesian  penalized likelihood approach.  Our frequentist approach and our calculation of standard errors  appear to be new.

\section{Parameter estimation via the EM algorithm}\label{EMsection}
In this section we describe how the EM algorithm can be used to obtain the parameter estimates for the penalized log-likelihood and Bayesian approaches. The E-step of the algorithm is exactly the same for both approaches. The M-step differs only in the part involving the calculation of $\hat{f}_j$. In the M-step we restrict our calculations to normally distributed errors. Furthermore, in the M-step, for the penalized log-likelihood case we can show that the maximizing $f_j$'s are cubic smoothing splines: our E-step leads to the maximization criterion for $f_j$ given by (\ref{S_line1}) plus (\ref{S_line2}), which is similar
to (5.1) in \citet{silverman1985some}. See also \citet{heckman2012}. We use this to justify modeling each $f_j$ as a linear combination of $K$ known B-spline basis functions $b_1,\ldots,b_K$, that is,
$$f_j(x)=\sum_{k=1}^K \phi_{jk} b_k(x),$$

\noindent with $\{\phi_{j1},\ldots,\phi_{jK}\}$ the set of unknown parameters determining $f_j$.

Recall that $\theta= \{ \alpha, \gamma \}$, where $\alpha$ is the vector containing the parameters of the model assumed for $\vect{z}$, and  $\gamma=\{f_j(\vect{x})\; \mbox{and} \;\sigma_j^2 \; \mbox{for} \; j=1,\ldots,J\}$ are the parameters governing the distribution of $y_i$ given the $f_j(\mathbf{x})$'s and the $z$'s. Let $\log p(\vect{y}|\theta)$ be the log-likelihood based on the observed data. Our goal is to find $\hat{\theta}$ that maximizes
\begin{equation}
 l(\theta) \equiv \log p(\vect{y}|\theta) + P(f_1,\ldots,f_J,\lambda_1,\ldots,\lambda_J).
\label{logposterior}
\end{equation}

The form of $\log p(\mathbf{y}|\theta)$ is very complicated, since it involves the distribution of the latent $z_i$'s. Therefore, the maximization of (\ref{logposterior}) with respect to $\theta$ is a difficult task. In order to tackle this problem it is common to apply numerical methods such as the EM algorithm to obtain the parameter estimates \citep{mclachlan}. The EM algorithm is usually used to maximize the likelihood function by generating a sequence of estimates, $\theta^{(c)}$, $c \geq 1$, having the property that  $\log p(\vect{y}|\theta^{(c+1)}) \geq  \log p(\vect{y}|\theta^{(c)})$. The EM algorithm can also be used to maximize (\ref{logposterior}). We can show (see \ref{suppA}) that our EM algorithm also generates a sequence of estimates, $\theta^{(c)}$, $c \geq 1$, satisfying
\begin{equation}
l(\theta^{(c+1)}) \geq l(\theta^{(c)}).
\label{ineq1}
\end{equation}

To define our EM algorithm, let $p(\vect{y},\vect{z}|\theta)$ be the joint distribution of the observed and latent data given $\theta$, also called the complete data distribution. In what follows view $\theta$ as an argument of a function, not as a random variable. Note that we write $\mbox{E}_{\theta^{(c)}}\big[H(\vect{y},\vect{z})|\mathbf{y}\big]$ to denote the conditional expected value of $H(\vect{y},\vect{z})$ assuming that the data $\vect{y}$ and $\vect{z}$ are generated with parameter vector $\theta^{(c)}$.

The proposed EM algorithm consists of the following two steps based on writing $\log p(\vect{y},\vect{z}|\theta) = \mathcal{L}_1(\gamma)+ \mathcal{L}_2(\alpha)$, where
\begin{equation*}
\mathcal{L}_1(\gamma) = \log p(\vect{y}|\vect{z},\theta) = \sum_{i=1}^n \log p(y_i|z_i,f_{z_i}(x_i),\sigma_{z_i}^2)
\end{equation*}
\noindent and
\begin{equation}
\label{L2general}
\mathcal{L}_2(\alpha) = \log p(\vect{z}| \theta) = \log p(z_1,\ldots, z_n| \alpha).\end{equation}

\begin{enumerate}
\item Expectation step (E-step): calculate
$$Q(\theta,\theta^{(c)}) \equiv \mbox{E}_{\theta^{(c)}}\big[\log p(\vect{y},\vect{z}|\theta)|\mathbf{y}\big] = \E_{\theta^{(c)}}(\el_1 (\gamma)| \mathbf{y}) + \E_{\theta^{(c)}}(\el_2 (\alpha)| \mathbf{y})$$
\noindent that is, calculate the
expected value of the logarithm of $p(\vect{y},\vect{z}|\theta)$ with respect to the distribution of the latent $\vect{z}$ given the observed $\vect{y}$ using $\theta^{(c)}$ as the true value of $\theta$.
\item Maximization step (M-step): let
$$S(\theta,\theta^{(c)})= Q(\theta,\theta^{(c)}) + P(f_1,\ldots,f_J,\lambda_1,\ldots,\lambda_J)$$
\begin{equation}
\label{Sfunction_Mstep}
= \E_{\theta^{(c)}}(\el_1 (\gamma)| \mathbf{y}) + P(f_1,\ldots,f_J,\lambda_1,\ldots,\lambda_J)+ \E_{\theta^{(c)}}(\el_2 (\alpha)| \mathbf{y}).
\end{equation}

\noindent Use the Expectation-Conditional Maximization (ECM) algorithm (see \ref{suppA}) to find $\theta^{(c+1)}$ that maximizes $S(\theta,\theta^{(c)})$ with respect to $\theta$ or at least does not decrease $S(\theta,\theta^{(c)})$ from the current value at $\theta^{(c)}$.
We can show (see \ref{suppA}) that if $S(\theta^{(c+1)},\theta^{(c)}) \geq S(\theta^{(c)},\theta^{(c)})$  then (\ref{ineq1}) holds.
\end{enumerate}

\subsection{E-step: general \texorpdfstring{$z_i$'s}{zs}}\label{generalzs}
\label{subsection:details}
Since
$$\el_1(\gamma) = \sum_{i=1}^n \sum_{j=1}^J \I ( z_i=j ) ~ \log p(y_i|z_i=j,f_{j}(x_i),\sigma_j^2),$$
\begin{equation}
\E_{\theta^{(c)}}(\el_1 (\gamma)| \mathbf{y}) =
\sum_{i=1}^n \sum_{j=1}^J p_{ij}^{(c)} \log p(y_i|z_i=j,f_{j}(x_i),\sigma_j^2) \nonumber
\label{Exp_EL1}
\end{equation}

\noindent where
\begin{equation*}
\pijc = p(z_i=j|\mathbf{y} , \theta^{(c)}),
\end{equation*}

\noindent whose exact form depends on the model for the $z_i$'s.

In a regression model with normal errors
\begin{eqnarray}
\label{normalerrors}
\E_{\theta^{(c)}}(\el_1 (\gamma)| \mathbf{y}) &=&  - ~ \frac{n}{2} \log (2 \pi) - \frac{1}{2}  \sum_{i=1}^n \sum_{j=1}^J p_{ij}^{(c)} \log \sigma_j^2 \\ \nonumber
&& - ~ \frac{1}{2}  \sum_{i=1}^n \sum_{j=1}^J p_{ij}^{(c)} \frac{[y_i - f_j(x_i)]^2}{ \sigma_j^2}.
 \end{eqnarray}

In the following sections we calculate $\E_{\theta^{(c)}}(\el_1(\gamma)|\mathbf{y})$ and $\E_{\theta^{(c)}}(\el_2(\alpha)|\mathbf{y})$ considering different models for the latent variables. Note that $\E_{\theta^{(c)}}(\el_1(\gamma)|\mathbf{y})$ depends on the model for the $z_i$'s only through $p_{ij}^{(c)}$.

\subsubsection{E-step: independent and identically distributed (\textit{iid}) \texorpdfstring{$z_i$'s}{zs}}
\label{subsection:independent}
In this section we calculate  $p_{ij}^{(c)}$ and $\E_{\theta^{(c)}}(\el_2(\alpha)|\mathbf{y})$ assuming that the latent variables $z_1,\ldots,z_n$ are \textit{iid} with parameter vector $\alpha=(p_1,...,p_J)$, where $p_j=p(z_i=j|\alpha)$ and $\sum_{j=1}^J p_j=1$.

Since the $z_i$'s are \textit{iid}, we obtain
\begin{equation}
p_{ij}^{(c)} = \frac{p(y_i|z_i=j,f_{j}(x_i)^{(c)},\sigma_{j}^{2\,(c)})\times p_j^{(c)}}{\sum_{l=1}^J  p(y_i|z_i=l,f_{l}(x_i)^{(c)},\sigma_{l}^{2\,(c)})\times p_l^{(c)}}.
\label{pij_iid}
\end{equation}

Note that we can easily calculate (\ref{pij_iid}) when the regression errors are normally distributed.

We now calculate $\E_{\theta^{(c)}}(\el_2(\alpha)|\mathbf{y})$. For \textit{iid} $z_i$'s we can write
\begin{equation}
\label{L2alphaiid}
 \el_2(\alpha) = \sum _{i=1}^n \sum_{j=1}^J \I(z_i=j)  \log p_j
\end{equation}

\noindent and, therefore,
\begin{equation}
\label{eq:EL2.independent}
\E_{\theta^{(c)}}(\el_2(\alpha)|\mathbf{y}) =  \sum_{i=1}^n \sum_{j=1}^J \pijc \log p_j.
\end{equation}

\subsubsection{E-step: Markov \texorpdfstring{$z_i$'s}{zs}}
\label{subsection:markov}
Here we calculate $\pijc$ and $\E_{\theta^{(c)}}(\el_2(\alpha)|\mathbf{y})$ assuming a Markov structure for the latent variables $z_1,\ldots,z_n$. In this case the distribution of the $z_i$'s depends on a vector $\alpha$ composed of transition probabilities and initial probabilities $\pi_j=p(z_1=j|\alpha)$, $j=1,\ldots,J$.

Let us assume that
\begin{enumerate}
\item  the $i$th latent variable depends on past latent variables only via the $(i-1)$st latent variable, i.e., $p(z_i | z_{i-1},\ldots,z_{1},\alpha)=p(z_i | z_{i-1},\alpha)$;
\item the transition probabilities do not depend on $i$, that is,
$$p(z_i=j|z_{i-1}=l,\alpha)=p(z_{i+s}= j | z_{i+s-1}=l,\alpha) \equiv a_{lj}.$$
\end{enumerate}

To compute $\pijc$ when the $z_i$'s are Markov, we use the results of \citet{baum1970}. These authors let
\begin{equation*}
\delta_{ij}^{(c)} =  p(y_1,\ldots,y_i,z_i=j | \theta^{(c)} )
\end{equation*}

\noindent and
\begin{equation*}
\varphi_{ij}^{(c)} = p(y_{i+1},\ldots,y_{n}|z_i=j,\theta^{(c)}),
\end{equation*}

\noindent and show how to calculate these recursively using what they call the forward and backward procedures, respectively.

Note that because of the Markovian conditional independence
\begin{eqnarray*}
\delta_{ij}^{(c)}\varphi_{ij}^{(c)}&=& p(y_1,\ldots,y_i,z_i=j | \theta^{(c)} ) \times p(y_{i+1},\ldots,y_{n}|z_i=j,\theta^{(c)})  \\
&=& p(y_1,\ldots,y_i,z_i=j | \theta^{(c)} ) \times p(y_{i+1},\ldots,y_{n}|z_i=j,y_1,\ldots,y_i, \theta^{(c)})  \\
&=& p( \mathbf{y},z_i=j | \theta^{(c)}).
\end{eqnarray*}

Thus, we can calculate $\pijc$ via
\begin{equation}
\pijc = \frac{\delta_{ij}^{(c)} \varphi_{ij}^{(c)}} {\sum_{l=1}^J\delta_{il}^{(c)} \varphi_{il}^{(c)}}.
\nonumber
\end{equation}

Now let us consider $\E_{\theta^{(c)}}(\el_2(\alpha)|\mathbf{y})$. Since for Markov $z_i$'s
\begin{equation}
\label{elalphamarkov}
\mathcal{L}_2(\alpha) = \sum_{i=2}^n \sum_{l=1}^J \sum_{j=1}^J \I(z_{i-1}=l, z_i=j) \log a_{lj}   + \sum_{j=1}^J \I(z_1=j) \log \pi_j
\end{equation}

\begin{equation}
\label{EL2.markov.final}
\E_{\theta^{(c)}}(\el_2(\alpha) |\mathbf{y}) =  \sum_{i=2}^n \sum_{l=1}^J \sum_{j=1}^J  p_{ilj}^{(c)} \,\log a_{lj}  + \sum_{j=1}^J p_{1j}^{(c)} \,\log \pi_{j},
\end{equation}

\noindent where $p_{ilj}^{(c)} \equiv p(z_{i-1}=l, z_i=j|\mathbf{y},\theta^{(c)})$. This can be expanded as
 $$p_{ilj}^{(c)} = \frac{p_{(i-1)l}^{(c)} \times a_{lj}^{(c)} \times p(y_{i}|z_{i}=j,\theta^{(c)}) \times \varphi_{ij}^{(c)}}{\varphi_{(i-1)l}^{(c)}},$$

\noindent which we easily calculate using the normality assumption for the regression errors.

More details on how to obtain the expressions for $\pijc$ and $p_{ilj}^{(c)}$ can be found in \citet{bilmes1998}, \citet{rabiner1989} and \citet{cappe2005}.

\subsection{M-step} \label{Mstep:details}
For the M-step we combine our discussion of \textit{iid} $z$'s and Markov $z$'s. We want to find $\theta^{(c+1)}$ that maximizes $S(\theta,\theta^{(c)})$ in (\ref{Sfunction_Mstep}) with respect to $\theta$ or at least produces a value of $S$ no smaller than $S(\theta^{(c)},\theta^{(c)})$. For normally distributed errors $\E_{\theta^{(c)}}(\el_1(\gamma) |\mathbf{y})$ is given by (\ref{normalerrors}) and, therefore, we can write
\begin{eqnarray}
S(\theta,\theta^{(c)})&=& \nonumber \\
&&  C - \frac{1}{2}  \sum_{i=1}^n \sum_{j=1}^J \pijc \log \sigma_j^2 - \frac{1}{2}  \sum_{i=1}^n \sum_{j=1}^J \pijc \frac{[y_i - f_j(x_i)]^2}{\sigma_j^2} \label{S_line1} \\
&&  + ~   P(f_1,\ldots,f_J,\lambda_1,\ldots,\lambda_J) \label{S_line2} \\
&& + ~ \E_{\theta^{(c)}}(\el_2(\alpha) |\mathbf{y}).
\label{S_line3}
\end{eqnarray}

We maximize $S$ as a function of $\theta=\{f_j(\vect{x}),\sigma_j^2,\, j=1,\ldots,J \; \mbox{and} \; \alpha\}$. In the maximization $\theta^{(c)}$ is fixed, not depending on $\theta$. This implies that the $\pijc$'s are also fixed, since their calculation depends on current parameter estimates in $\theta^{(c)}$. We also consider the smoothing parameters, $\lambda_1,\ldots,\lambda_J$, to be fixed. This maximization cannot be done analytically, so we apply a natural extension of the EM approach, the ECM algorithm, to guarantee that $S(\theta^{(c+1)},\theta^{(c)}) \geq S(\theta^{(c)},\theta^{(c)})$.

Because the expression for $\E_{\theta^{(c)}}(\el_2(\alpha) |\mathbf{y})$ does not depend on the $f_j$'s or $\sigma_j^2$'s, the $f_j$'s and  $\sigma_j^2$'s that maximize $S$ are the $f_j$'s and  $\sigma_j^2$'s that maximize (\ref{S_line1}) + (\ref{S_line2}). Therefore, the form of the  maximizing $f_j$'s and  $\sigma_j^2$'s  will not depend on the model for the $z$'s. Their only dependence on the model for the $z$'s is via the $p_{ij}^{(c)}$'s.

Thus, to obtain the vector of parameter estimates, $\theta^{(c+1)}$, we apply the ECM algorithm as follows.
\begin{enumerate}
\item Hold the $\sigma^2_j$'s and the parameters in $\alpha$ fixed and maximize (\ref{S_line1}) plus (\ref{S_line2}) with respect to $f_j(\vect{x})$. Let
\begin{equation}
\label{Wjdef}
\matr{W}_j=\mbox{diag}(p_{1j}^{(c)}/\sigma^2_j,\ldots,p_{nj}^{(c)}/\sigma^2_j).
\end{equation}

\noindent For the Bayesian approach this is equivalent to maximizing
$$  - \frac{1}{2} \sum_{j=1}^J \big(\vect{y}-f_j(\vect{x})\big)^T\matr{W}_j\big(\vect{y}-f_j(\vect{x})\big) - \frac{1}{2} \sum_{j=1}^J  f_j(\vect{x})^T ~ \Covfj^{-1} ~f_j(\vect{x})  $$

\noindent  obtaining
$$\hat{f}_j(\vect{x})=  \Covfj \big( \Covfj + \matr{W}_j^{-1} \big )^{-1}  {\bf{y}}.$$

\noindent Let $f_j^{(c+1)}(\vect{x})$ be $\hat{f}_j(\vect{x})$ with $\sigma^2_j$ in $\matr{W}_j$  replaced by  $\sigma^{2\,(c)}_j$.
\\

For the penalized log-likelihood approach recall that $f_j$ is a linear combination of $K$ known basis functions, so $f_j(\vect{x})=\matr{B}\phi_j$, where $\phi_j=(\phi_{j1},\ldots,\phi_{jK})^T$ is the vector of coefficients corresponding to $f_j$ and $\matr{B}$ is an $n\times K$ matrix with entries $B_{ik}=b_k(x_i)$. Thus, we hold the $\sigma_j^2$'s and $\alpha$ fixed and maximize
\begin{equation*}
  - \frac{1}{2}  \sum_{j=1}^J
(\vect{y}-\matr{B}\phi_j)^T \matr{W}_j(\vect{y}-\matr{B}\phi_j) - \sum_{j=1}^J \lambda_j \phi_j^T \matr{R} \phi_j,
\end{equation*}

\noindent with respect to $\phi_j$ yielding
$$\hat{\phi}_j = (\matr{B}^T\matr{W}_j\matr{B} + 2\lambda_j\matr{R})^{-1}\matr{B}^T\matr{W}_j\vect{y},$$

\noindent where $\matr{R}$ is a $K \times K$  matrix with entries
$$R_{kk'}=\int b''_k(x) b''_{k'}(x)~dx.$$

Let $\phi_j^{(c+1)}$ be $\hat{\phi}_j$ with $\sigma_j^2$ in $\matr{W}_j$ replaced by $\sigma_j^{2(c)}$. So we let $f_j^{(c+1)}(\vect{x})=\matr{B}\phi_j^{(c+1)}$.
\\

\item Now holding the $f_j(\vect{x})$'s and the parameters in $\alpha$ fixed and maximizing (\ref{S_line1}) with respect to $\sigma^2_j$ we get
\begin{equation}
\hat{\sigma}_j^2 = \frac{\displaystyle \sum_{i=1}^n  p^{(c)}_{ij} \big[ y_i - f_j(x_i)\big]^2} {\displaystyle  \sum_{i=1}^n p^{(c)}_{ij}}.
\label{difsigma}
\end{equation}

\noindent Let $\sigma_j^{2\,(c+1)}$ be $\hat{\sigma}_j^2$ with $f_j(x_i)$ replaced by $f_j(x_i)^{(c+1)}$. If we assume $\sigma^2_j=\sigma^2$ for all $j$, then we find that
\begin{equation}
\hat{\sigma}^{2\,(c+1)}=\frac{1}{n}\displaystyle \sum_{j=1}^J \sum_{i=1}^n p^{(c)}_{ij} \big[ y_i - f_j(x_i)^{(c+1)}\big]^2.
\label{samesigma}
\end{equation}

We can obtain better variance estimates by adjusting the degrees of freedom to account for the estimation of the $f_j$'s. We do that by replacing the denominator of (\ref{difsigma}) by  $\sum_{i=1}^n p^{(c)}_{ij} - \mbox{trace}(\matr{D}_j\matr{H}_j)$ and the denominator of (\ref{samesigma}) by $n - \sum_{j=1}^J\mbox{trace}(\matr{D}_j\matr{H}_j)$, where $\matr{D}_j=\mbox{diag}(p_{1j},\ldots,p_{nj})$ and $\matr{H}_j$ is the so called hat matrix satisfying $\hat{f}_j(\vect{x})=\matr{H}_j\vect{y}$. For the Bayesian approach $\matr{H}_j = \Covfj ( \Covfj + \matr{W}_j^{-1})^{-1}$ and for the penalized approach
$\matr{H}_j = \matr{B}(\matr{B}^T\matr{W}_j\matr{B} + 2\lambda_j\matr{R})^{-1}\matr{B}^T\matr{W}_j$. This modification is similar to a weighted version of what \citet{wahba1983bayesian} has proposed for the regular smoothing spline case.
\\

\item Now we hold the $f_j(\vect{x})$'s and $\sigma_j^2$'s fixed and maximize $S$ with respect to the parameters in $\alpha$. Note that
(\ref{S_line1}) and (\ref{S_line2}) do not depend on $\alpha$, so to find $\alpha^{(c+1)}$, we maximize $\E_{\theta^{(c)}}(\el_2(\alpha) |\mathbf{y})$ in line (\ref{S_line3}) as a function of $\alpha$. Because the form of $\E_{\theta^{(c)}}(\el_2(\alpha) |\mathbf{y})$ does depend on the model for the $z$'s, we must obtain the estimates of $\alpha$ for each model separately.
\\

For \textit{iid} $z_i$'s, where $p_j=p(z_i=j|\alpha)$, using (\ref{eq:EL2.independent}) and Lagrange multipliers with the restriction that $\sum_{j=1}^J p_j=1$, we obtain:
$$p^{(c+1)}_j = \frac{1}{n}\sum_{i=1}^n p^{(c)}_{ij}.$$

For Markov $z$'s the vector $\alpha$ is composed of transition probabilities $a_{lj}$ and initial probabilities $\pi_j$. We first maximize $\E_{\theta^{(c)}}(\el_2(\alpha) |\mathbf{y})$ given in (\ref{EL2.markov.final}) with respect to $a_{lj}$. Holding $\pi_j$ fixed and using a Lagrange multiplier with the constraint $\sum_{j=1}^J a_{lj}=1$, we get:
$$a_{lj}^{(c+1)}=  \frac{\displaystyle \sum_{i=2}^n p_{ilj}^{(c)}}{\displaystyle \sum_{i=2}^n p_{(i-1)l}^{(c)}}.$$

Now let us maximize (\ref{EL2.markov.final}) with respect to $\pi_j$. Holding $a_{lj}$ fixed and using a Lagrange multiplier with the restriction that $\sum_{j=1}^J \pi_j=1$, we obtain:
$$\pi_j^{(c+1)} = p^{(c)}_{1j}.$$
\end{enumerate}

\section{Standard errors for the parameter estimators of the state process} \label{sec:standarderrors}

In this section we use the results of \citet{louis1982} to obtain standard errors for the estimates of the parameters of the state process. We consider for \textit{iid} $z_i$'s $J\geq2$ possible state values. For Markov $z_i$'s we restrict the possible number of states to $J=2$ to reduce calculational complexity.

\citet{louis1982} derived a procedure to obtain the observed information matrix when the maximum likelihood estimates are obtained using the EM algorithm. The procedure requires the computation of the gradient and of the second derivative matrix of the log-likelihood based on the complete data and can be implemented quite easily within the EM steps.

\subsection{The general case}

Suppose that $\gamma$ is known and let $\hat{\alpha}^*=\hat{\alpha}^*(\gamma)$ be the maximum likelihood estimator of $\alpha$ given $\gamma$, that is, the maximizer of $L_I(\alpha)\equiv \log p(\vect{y}|\alpha)$, the incomplete data log-likelihood.
We can obtain $\hat{\alpha}^*$ using the EM algorithm, and the complete data log-likelihood $L_C(\alpha) \equiv \log p(\vect{y},\vect{z}|\alpha)= \mathcal{L}_2(\alpha) + C $ (see Section \ref{generalzs}) . In this case we can derive the observed information matrix, $I_{\gamma}(\alpha)$, via a direct application of Louis's procedure. Under some regularity conditions, \citet{louis1982} shows by straightforward differentiation that $L_I'(\alpha)=E_{\alpha}(L_C'(\alpha)|\vect{y})$, $L_I'(\hat{\alpha}^*)= 0$, and that the observed information matrix is given by
\begin{eqnarray}
\label{information}
I_{\gamma}(\alpha) &=&  - L_I''(\alpha) \nonumber \\
  &=&E_{\alpha}(- \mathcal{L}_2''(\alpha)|\vect{y}) - E_{\alpha}( \mathcal{L}_2'(\alpha) \mathcal{L}_2'(\alpha)^T|\vect{y}) + L_I'(\alpha)L_I'(\alpha)^T,
\end{eqnarray}

\noindent where $\mathcal{L}_2(\alpha)$ is as in (\ref{L2general}), $\mathcal{L}'_2(\alpha)$ and $L'_I(\alpha)$ are the gradient vectors of $\mathcal{L}_2$ and $L_I$, respectively, and $\mathcal{L}''_2(\alpha)$ and $L_I''(\alpha)$ are the associated second derivative matrices. The estimate of $I_{\gamma}(\alpha)$ is $I_{\gamma}(\hat{\alpha}^*)$. Note that (\ref{information}) needs to be evaluated only at convergence of the EM algorithm, where $L_I'$ is zero.  Then, $I_{\gamma}(\hat{\alpha}^*)=I_{\gamma}(\hat{\alpha}^*(\gamma))$ contains only the first two terms of (\ref{information}). The inverse of $I_{\gamma}(\hat{\alpha}^*)$ is the estimated variance-covariance matrix of $\hat{\alpha}^*$ for the known value of $\gamma$.

To relate these calculations to those of our EM algorithm, where both $\alpha$ and $\gamma$ are estimated, first note that our $\hat{\gamma}$ and $\hat{\alpha}$ are the maximizers of the criterion $ \log p(\vect{y}|\gamma,\alpha) + P(f_1,\ldots,f_J,\lambda_1,\ldots,\lambda_J)$. If the maximizer is unique then $\hat{\alpha}=\hat{\alpha}^*({\hat{\gamma}})$. Therefore, we will estimate the variance-covariance matrix of $\hat{\alpha}$ by the variance-covariance matrix of $\hat{\alpha}^*$ plugging in $\gamma=\hat{\gamma}$. That is, we propose to use a plug-in estimate, $I_{\hat{\gamma}}(\hat{\alpha})$, where $\hat{\alpha}$ and $\hat{\gamma}$ are obtained from our EM procedure. Note that this method ignores the variability in estimating $\gamma$.

In the next sections we show how to calculate $I_{\hat{\gamma}}(\hat{\alpha})$ for the different models for the $z_i$'s.

\subsection{Standard errors: \textit{iid} \texorpdfstring{$z_i$'s}{zs}}
To remove the restriction $\sum_{j=1}^J{p_j} =1$, we use the parameters $p_1,\ldots,p_{J-1}$ and rewrite (\ref{L2alphaiid}) as
\begin{equation}
\mathcal{L}_2(\alpha) = \sum_{i=1}^n \left\{ \sum_{j=1}^{J-1} \I(z_i=j) \log p_j   +   \I(z_i=J) \log \Big(1-\sum_{j=1}^{J-1} p_j \Big)  \right \}.
\label{newL2alpha}
\end{equation}

\noindent Let $\mathcal{L}_2'(\alpha)$ be the $(J-1) \times 1$ gradient vector of (\ref{newL2alpha}) with the $j$th component given by

$$\frac{\partial \mathcal{L}_2}{\partial p_j} = \sum_{i=1}^n \left [ \frac{\I(z_i=j)}{p_j} - \frac{\I(z_i=J)}{(1-\sum_{k=1}^{J-1} p_k)}\right]$$

\noindent and $\mathcal{L}''_2(\alpha)$ be the $(J-1)\times (J-1)$ matrix with the associated second derivatives
$$\frac{\partial^2 \mathcal{L}_2}{\partial p_j^2} = - \sum_{i=1}^n \left [ \frac{\I(z_i=j)}{p_j^2} + \frac{\I(z_i=J)}{(1-\sum_{k=1}^{J-1} p_k)^2}\right],$$

$$\frac{\partial^2 \mathcal{L}_2}{\partial p_jp_l} = - \sum_{i=1}^n \left [  \frac{ \I(z_i=J)}{(1-\sum_{k=1}^{J-1} p_k)^2}\right] ~ \mbox{for} ~ j\neq l.$$

Consider the $(J-1)\times (J-1)$ matrix $E_{\alpha}(-\mathcal{L}''_2(\alpha)|\vect{y})$ in (\ref{information}) evaluated at $\alpha=\hat{\alpha}$. One can show its $jl$th entry, for $j \neq l$, is equal to $n/(1-\sum_{k=1}^{J-1}\hat{p}_k)$ and its $jj$th entry is
$$ \sum_{i=1}^n
\left( \frac{\hat{p}_{ij}}{\hat{p}_j^2} +  \frac{(1-\sum_{k=1}^{J-1}\hat{p}_{ik})}{(1-\sum_{k=1}^{J-1}\hat{p}_k)^2} \right)
= n \times \left(\frac{1}{\hat{p}_j} + \frac{1}{1- \sum_{k=1}^{J-1}\hat{p}_k}\right) $$

\noindent Note that the simplification above is obtained using the fact that $\hat{p}_j=\sum_{i=1}^n \hat{p}_{ij} / n$.

One can also show that the $(J-1)\times (J-1)$ matrix $E_{\alpha}(\mathcal{L}'_2(\alpha)\mathcal{L}'_2(\alpha)^T|\vect{y})$ in (\ref{information}) evaluated at $\hat{\alpha}$ has off diagonal elements $jl$ equal to
$$\frac{n}{(1-\sum_{k=1}^{J-1}\hat{p}_k)}
  - \sum_{i=1}^n \left[ \left( \frac{\hat{p}_{ij}}{\hat{p}_j} -  \frac{(1-\sum_{k=1}^{J-1}\hat{p}_{ik})}{(1-\sum_{k=1}^{J-1}\hat{p}_k)} \right)
\times \left( \frac{\hat{p}_{il}}{\hat{p}_l} -  \frac{(1-\sum_{k=1}^{J-1}\hat{p}_{ik})}{(1-\sum_{k=1}^{J-1}\hat{p}_k)} \right) \right]
$$

\noindent and $j$th diagonal element given by
$$  n \times \left(\frac{1}{\hat{p}_j} + \frac{1}{1- \sum_{k=1}^{J-1}\hat{p}_k}\right)
- \sum_{i=1}^n  \left( \frac{\hat{p}_{ij}}{\hat{p}_j} -  \frac{(1-\sum_{k=1}^{J-1}\hat{p}_{ik})}{(1-\sum_{k=1}^{J-1}\hat{p}_k)} \right)^2. $$

\subsection{Standard errors: Markov \texorpdfstring{$z_i$'s}{zs}}\label{seMarkov}

For Markov $z_i$'s we show how to obtain standard errors for the estimates of the transition probabilities for $J=2$ possible state values.
We apply Louis's method to find standard errors for $\hat{a}_{12}$ and $\hat{a}_{21}$ by first considering $\mathcal{L}_2$ in (\ref{elalphamarkov}) with $\pi_1$ and $\pi_2$ fixed. Abusing notation slightly by omitting $\pi_1$ and $\pi_2$, we let $a_{11} = 1-a_{12}$ and $a_{22}=1-a_{21}$ and write
\begin{eqnarray}
\mathcal{L}_2(a_{12},a_{21})&=& \sum_{i=2}^n \Big[   \I(z_{i-1}=1,z_{i}=2) \log a_{12}  \nonumber \\
&& ~~~~~ + ~ \I(z_{i-1}=1,z_{i}=1) \log (1-a_{12})   \Big. \nonumber \\
&& ~~~~~ + ~ \Big. \I(z_{i-1}=2,z_{i}=1) \log a_{21}  \nonumber  \\
&& ~~~~~ + ~   \I(z_{i-1}=2,z_{i}=2) \log (1-a_{21})   \Big]. \nonumber
\end{eqnarray}

The required two dimensional gradient vector is given by
$$   \left( \begin{array}{c}
\displaystyle\frac{\partial \mathcal{L}_2}{\partial a_{12}}  \\ [\bigskipamount] 
\displaystyle\frac{\partial \mathcal{L}_2}{\partial a_{21}}
 \end{array} \right) = \left( \begin{array}{c}
\displaystyle\sum_{i=2}^n \left [ \frac{\I(z_{i-1}=1,z_{i}=2)}{a_{12}} - \frac{(\I(z_{i-1}=1,z_{i}=1)}{(1-a_{12})}\right]  \\ [\bigskipamount]
\displaystyle\sum_{i=2}^n \left [ \frac{\I(z_{i-1}=2,z_{i}=1)}{a_{21}} - \frac{\I(z_{i-1}=2,z_{i}=2)}{(1-a_{21})}\right]
 \end{array} \right).  $$

The associated $2 \times 2$ matrix of second derivatives is diagonal with entries
$$\frac{\partial^2 \mathcal{L}_2}{\partial a_{12}^2} = - \sum_{i=2}^n \left [ \frac{\I(z_{i-1}=1,z_{i}=2)}{a_{12}^2} + \frac{\I(z_{i-1}=1,z_{i}=1)}{(1-a_{12})^2}\right]$$
\noindent and
$$\frac{\partial^2 \mathcal{L}_2}{\partial a_{21}^2} = - \sum_{i=2}^n \left [ \frac{\I(z_{i-1}=2,z_{i}=1)}{a_{21}^2} + \frac{\I(z_{i-1}=2,z_{i}=2)}{(1-a_{21})^2}\right].$$

Thus the $2 \times 2$ matrix $E_{\alpha}(-\mathcal{L}''_2(a_{12},a_{21})|\vect{y})$ with $\pi_1$ and $\pi_2$ fixed, evaluated at $a_{12}=\hat{a}_{12}$, $a_{21}=\hat{a}_{21}$ is given by
\[ \left( \begin{array}{cc}
\displaystyle \frac{\sum_{i=2}^n \hat{p}_{(i-1)1}}{\hat{a}_{12}(1-\hat{a}_{12})} & 0  \\
0 & \displaystyle \frac{ \sum_{i=2}^n \hat{p}_{(i-1)2}}{\hat{a}_{21}(1-\hat{a}_{21})}
 \end{array} \right).\]

Calculating the $2 \times 2$ matrix $E_{\alpha}(\mathcal{L}'_2(a_{12},a_{21})\mathcal{L}'_2(a_{12},a_{21})^T|\vect{y})$ is straightforward but tedious, involving sums of expectations of indicator functions. The summands require the calculation of $p(z_{i-1}=r,z_{i}=s|\vect{y},\alpha)$ , $p(z_{i-2}=r,z_{i-1}=s,z_{i}=t|\vect{y},\alpha)$ and $p(z_{i-1}=r,z_{i}=s,z_{i+\Delta}=t,z_{i+\Delta+1}=u|\vect{y},\alpha)$, with $r,s,t$ and $u$ taking values 1 or 2 and $\Delta$ a positive integer.  These conditional probabilities can be calculated using  Bayes' Theorem and the Markovian conditional independence of the $z_i$'s.

\section{Simulations}\label{sec:implem}

We carry out simulation studies considering that the $z_i$'s can take values 1 or 2 and they can be either $iid$ or follow a Markov structure. The parameters of interest are estimated using both the Bayesian and the penalized log-likelihood approaches presented in Section \ref{EMsection}. For each simulation study 300 independent data sets are generated.

\subsection{Simulated data}

We consider three types of simulation studies according to three different types of simulated data. Table \ref{tb:sim:summary} presents a summary of these simulation studies. In all studies we use the same vector of evaluation points $\vect{x}$ and the same true functions $f_1$ and $f_2$. The vector $\vect{x}=(x_1,\ldots,x_n)^T$ consists of $n=199$ equally spaced points, $1, 1.5, \ldots, 99.5, 100$. The true functions evaluated at $\vect{x}$, that is, the vectors $f_1(\vect{x})$ and $f_2(\vect{x})$, are obtained by sampling from a multivariate normal distribution with mean of zero and covariance matrix determined by (\ref{cov}). We let the parameter $U_j$ in (\ref{cov}) be equal to $1/\big(s_j \sqrt{2\pi}\big)$ so that $\lambda_j=s_j$. We consider $\lambda_1=s_1=28$ and $\lambda_2=s_2=38$ for $f_1(\vect{x})$ and $f_2(\vect{x})$, respectively.

After we obtain the commonly used $f_1(\vect{x})$ and $f_2(\vect{x})$ we generate a set of simulated data as follows.
\begin{enumerate}
\item Generate the $z_i$'s according to the specified model (\textit{iid} or Markov).
\item Generate the $y_i$'s with common regression error variance $\sigma^2$ as follows:
  \begin{itemize}
  \item if $z_i = 1$,  $y_i = f_1(x_i) + \sigma \epsilon_i$;
  \item if $z_i = 2$,  $y_i = f_2(x_i) + \sigma \epsilon_i $,
  \end{itemize}
  \noindent where $\sigma = 5\times 10^{-5}$ and $\epsilon_i$ has a $N(0,1)$ distribution.
\item Repeat steps 1 and 2 $S$ times obtaining $S$ different data sets.
\end{enumerate}

\noindent In our case $S=300$. Note again that $f_1(\vect{x})$ and $f_2(\vect{x})$ are fixed across all data sets.

\begin{figure}
        \centering
        \begin{subfigure}[b]{0.5\textwidth}
                \centering
                \includegraphics[width=\textwidth]{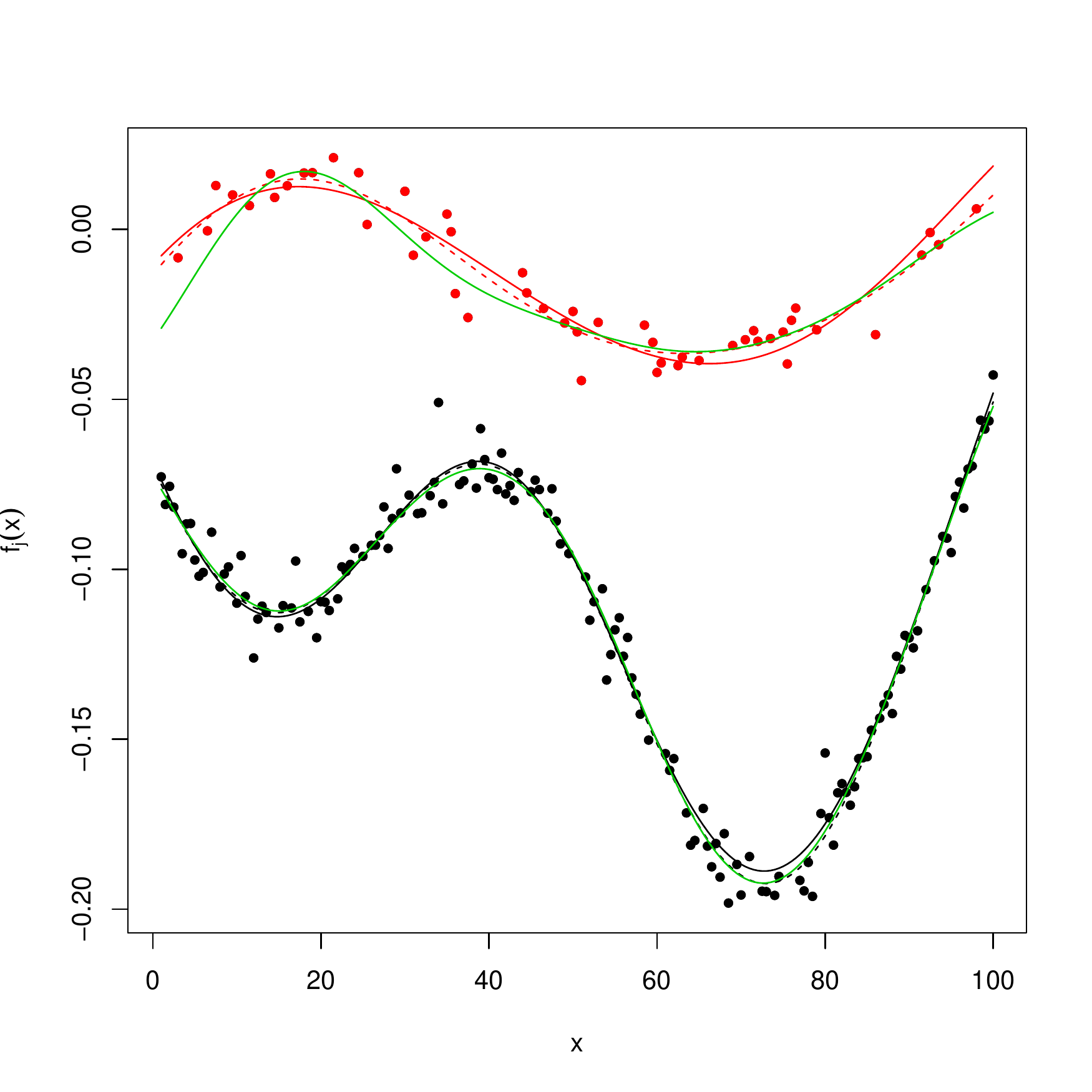}
                \caption{Bayesian}
        \end{subfigure}%
        ~ 
        \begin{subfigure}[b]{0.5\textwidth}
                \centering
                \includegraphics[width=\textwidth]{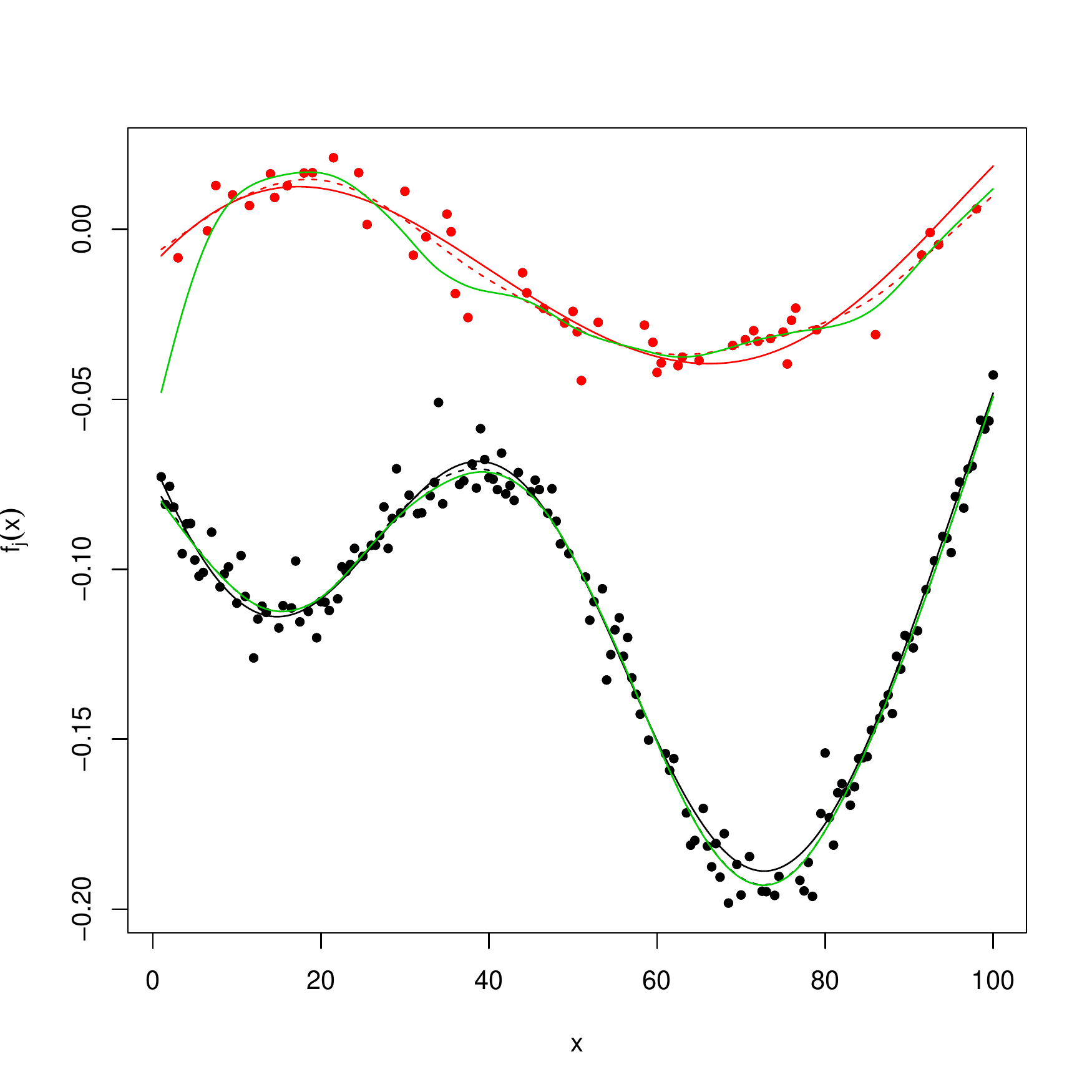}
                \caption{Penalized log-likelihood}
        \end{subfigure}
        ~ 
\caption{Simulation 1 (\textit{iid} $z_i$'s). Example of simulated data along with the initial and final estimates of $f_1$ and $f_2$ obtained using in (a) the Bayesian approach  and in (b) the penalized log-likelihood approach. The red dots correspond to $z=2$  and the black dots to $z=1$. The solid lines correspond to the true functions $f_1$ and $f_2$. The green lines are the initial estimates of $f_1$ and $f_2$ obtained using the function estimate method. The dashed lines are the final estimates of $f_1$ and $f_2$.}
\label{S1Curves}
\end{figure}

In our first simulation study we generate data assuming that the $z_i$'s are independent with $p_1=p(z_1=1)=0.7$. Figure \ref{S1Curves} shows an example of a data set of this type.

\begin{figure}
        \centering
        \begin{subfigure}[b]{0.5\textwidth}
                \centering
                \includegraphics[width=\textwidth]{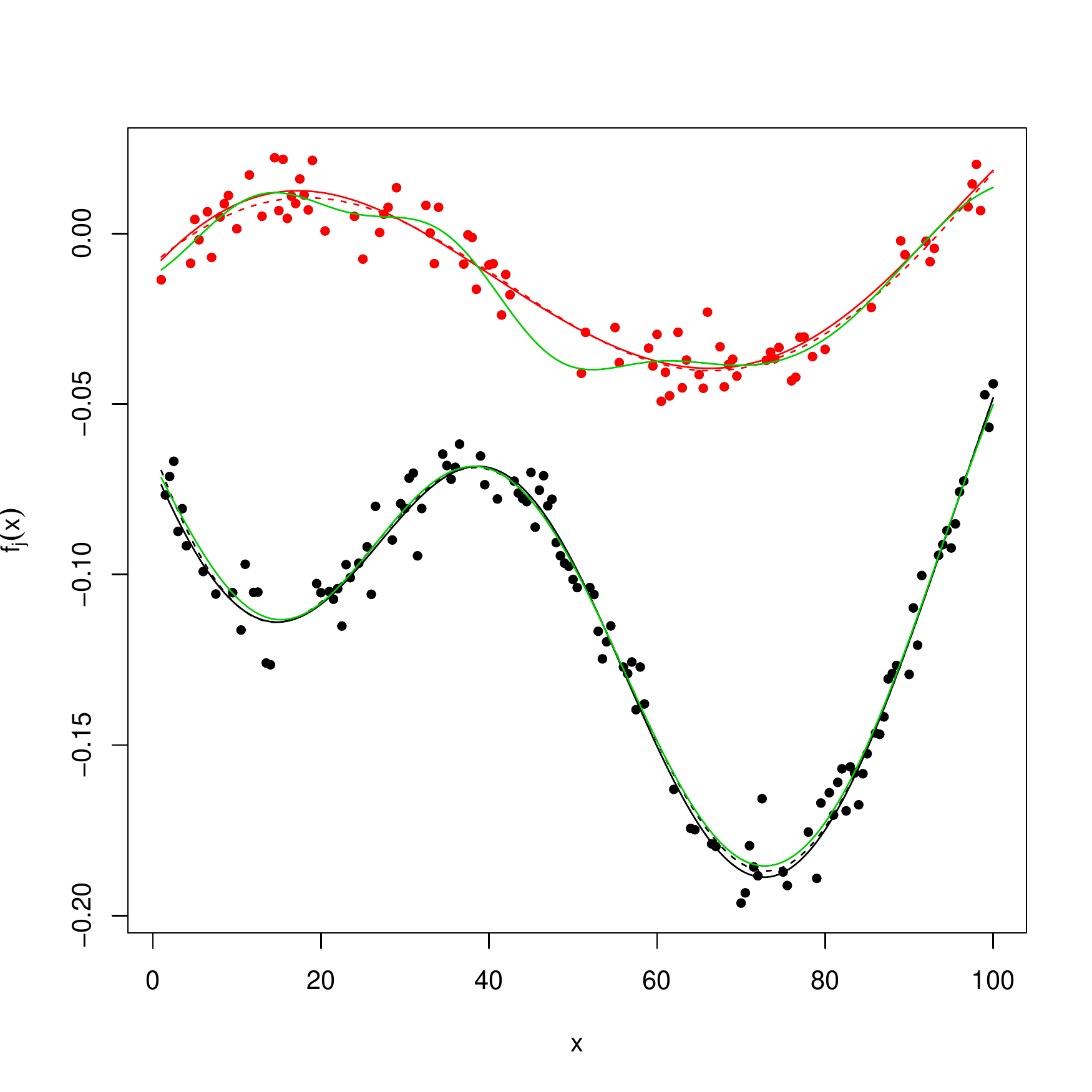}
                \caption{Bayesian}
        \end{subfigure}%
        \begin{subfigure}[b]{0.5\textwidth}
                \centering
                \includegraphics[width=\textwidth]{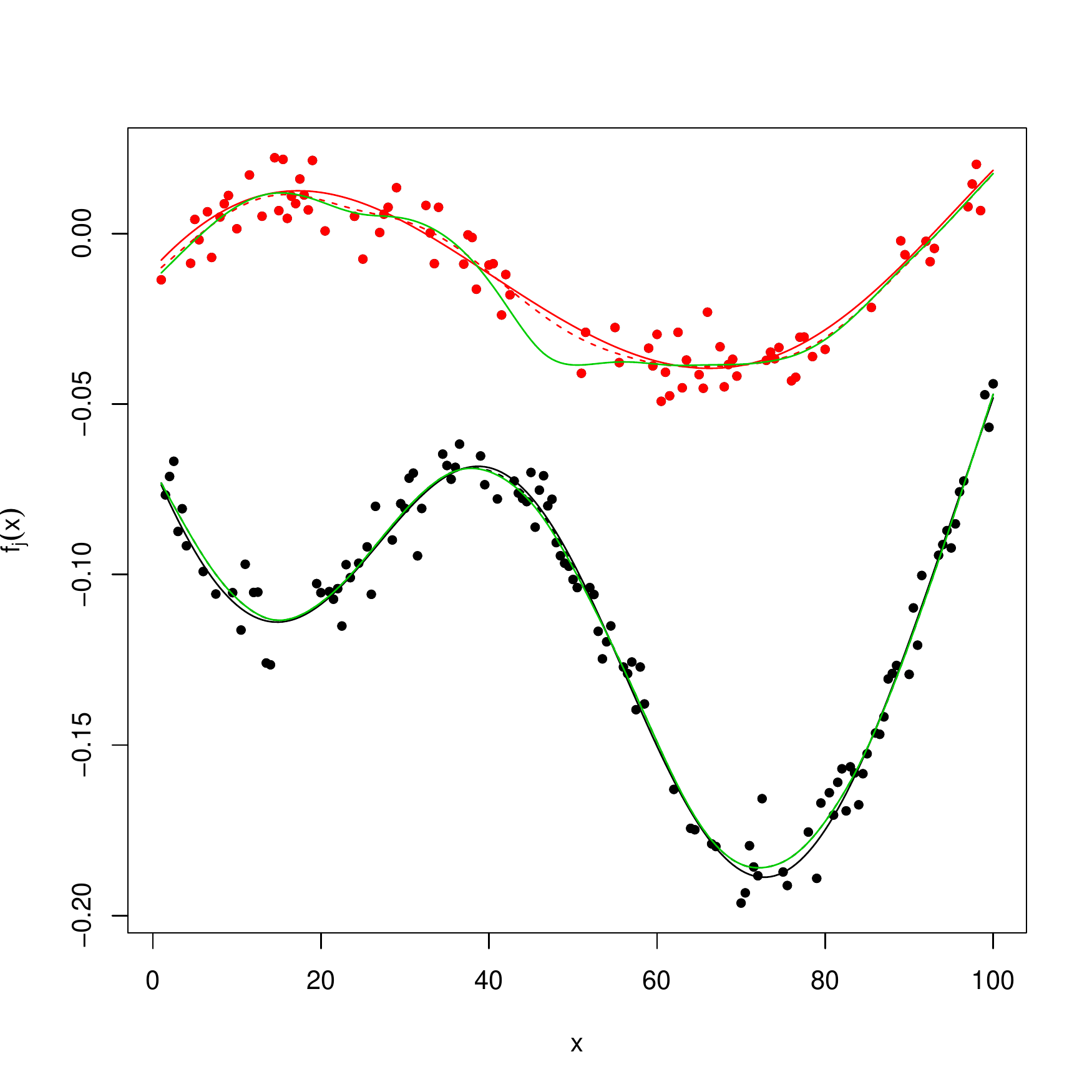}
                \caption{Penalized log-likelihood}
        \end{subfigure}
\caption{Simulation 2 (Markov $z_i$'s with $a_{12}=0.3$ and $a_{21}=0.4$). Example of simulated data along with the initial and final estimates of $f_1$ and $f_2$ as in Figure \ref{S1Curves}.}
\label{S2Curves}
\end{figure}

\begin{figure}
        \centering
        \begin{subfigure}[b]{0.5\textwidth}
                \centering
                \includegraphics[width=\textwidth]{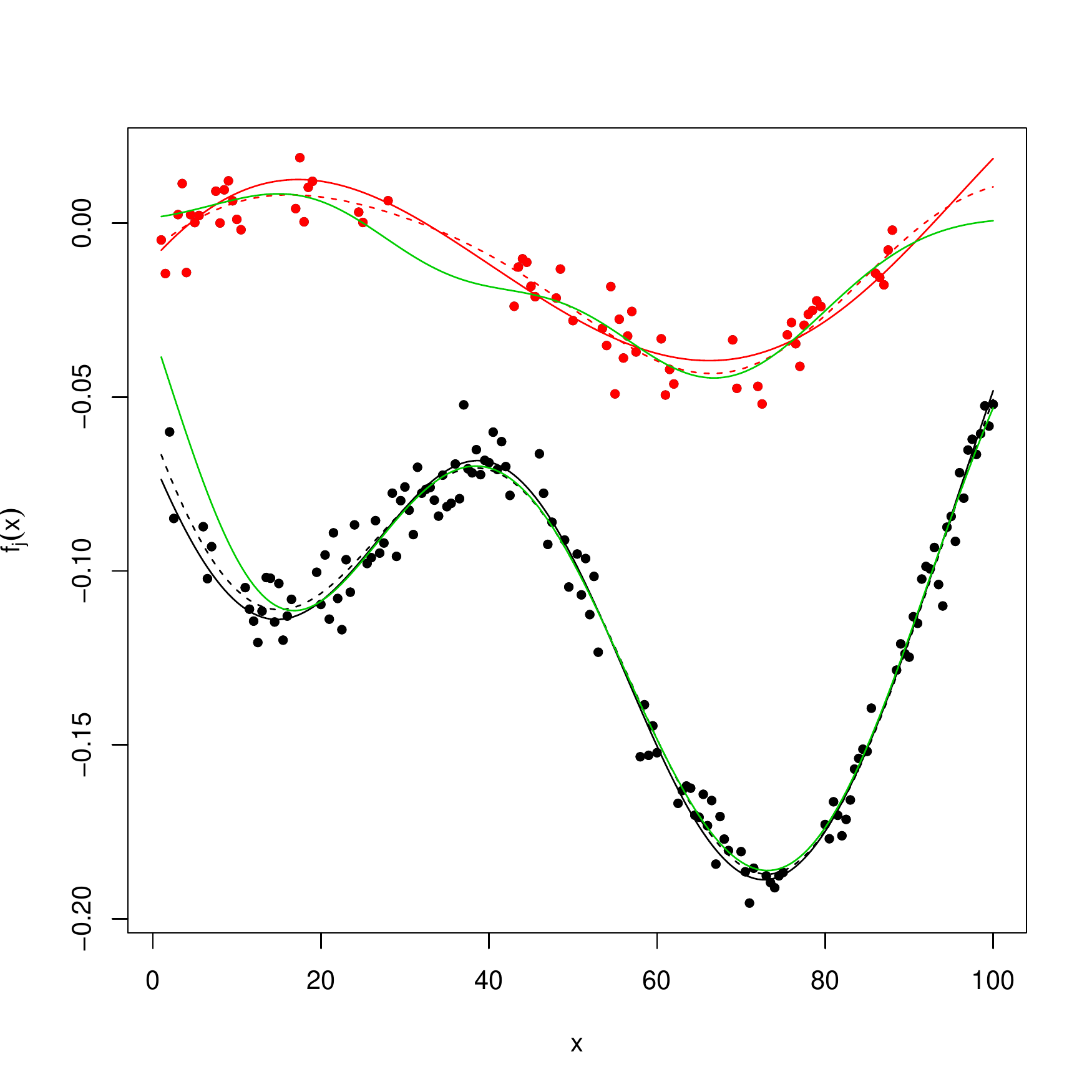}
                \caption{Bayesian}
        \end{subfigure}%
        \begin{subfigure}[b]{0.5\textwidth}
                \centering
                \includegraphics[width=\textwidth]{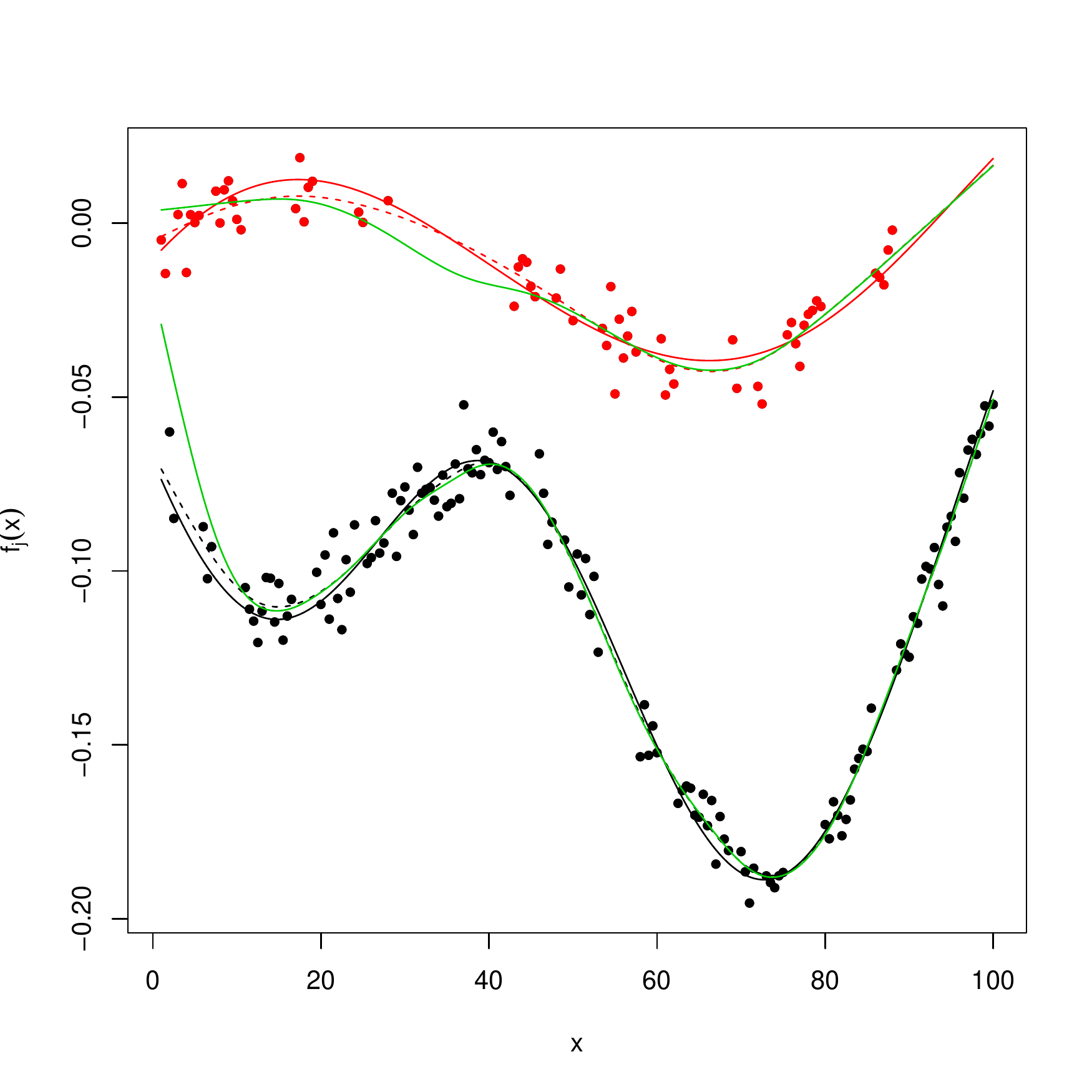}
                \caption{Penalized log-likelihood}
        \end{subfigure}
\caption{Simulation 3 (Markov $z_i$'s with $a_{12}=0.1$ and $a_{21}=0.2$). Example of simulated data along with the initial and final estimates of $f_1$ and $f_2$ as in Figure \ref{S1Curves}.}
\label{S3Curves}
\end{figure}

For the second and third simulation studies we consider Markov $z_i$'s. In the second study we use transition probabilities  $a_{12}=p(z_i=2|z_{i-1}=1)=0.3$ and $a_{21}=p(z_i=1|z_{i-1}=2)=0.4$ and in the third study $a_{12}=0.1$ and $a_{21}=0.2$. For both studies we consider initial probabilities $\pi_1=\pi_2=0.5$. Figures \ref{S2Curves} and \ref{S3Curves} show examples of data sets from the second and third studies, respectively. Compared to Figure \ref{S1Curves}, in Figures \ref{S2Curves} and \ref{S3Curves} we observe that the system can stay in one state for a long time, giving information about just one of the functions during that range of $x$ values. This is more pronounced in Figure \ref{S3Curves} when $a_{12}$ and $a_{21}$ are small.

\subsection{Initial values}

To analyze the data via our EM algorithm, we need to provide initial values of all of the parameter estimates.

We set the initial estimates of all of the parameters governing the distribution of the $z_i$'s to 0.5.

In order to automatically obtain initial values for $f_1(\vect{x})$ and $f_2(\vect{x})$ we first assign temporary values to the latent variables, creating two groups of observations (one consisting of all $(x_i,y_i)$'s with temporary $z_i$ value equal to 1, the other consisting of the remaining $(x_i,y_i)$'s). For each group, we estimate the corresponding $f_j(\vect{x})$. For the Bayesian approach we estimate $f_j(\vect{x})$ by its posterior mean, commonly used in Gaussian regression \citep{rasmussen2006}, with initial $\sigma^2=0.005$. For the penalized log-likelihood approach we use smoothing splines to estimate the $f_j$'s. The smoothing parameters for both approaches are chosen by generalized cross-validation (GCV). For the Bayesian approach we also add the restriction that if the $\lambda_j$ obtained by GCV is smaller than 10 we use a bigger value, in this case 15. Two methods are used to assign the temporary values to the $z_i$'s.
\begin{itemize}
\item \textit{Function estimate method:} we fit one curve,  $\hat{m}(\cdot)$, to the whole data set using a standard cubic smoothing spline. If $y_i \leq \hat{m}(x_i)$, we set $z_i=1$, otherwise we set $z_i=2$.
\item \textit{Residual-based method:} we obtain $\hat{m}(\cdot)$ as in the function estimate method and calculate the residuals $y_i - \hat{m}(x_i)$. We then divide the evaluation interval into small subintervals. Within each subinterval, we consider all residuals corresponding to $x_i$'s within that subinterval. We use the $k$-means algorithm ($k=2$) to partition these residuals into two groups. We label the group with the smaller mean as group 1 $(z_i\mbox{'s}=1)$.  The $k$-means algorithm is a clustering method that aims to partition observations into $k$ groups such that the sum of the squared differences between each observation and its assigned group mean is minimized \citep{johnson2008}.
\end{itemize}

The function estimate method is used in Simulations 1 and 2, and the residual-based method in Simulation 3. The green lines in Figures \ref{S1Curves}, \ref{S2Curves} and \ref{S3Curves} are examples of initial functions.

In Simulation 3, we use the residual-based method because the probabilities of changing from one state to another are small, that is, the $z$ process tends to remain in one state for an extended period, as shown in Figure \ref{S3Curves}. During this period, we only obtain information from one of the $f_j$'s and in this case the function estimate method fails to produce good initial values. The residual-based method requires a choice of number of sub-intervals and sub-interval lengths. Our choice of sub-intervals is based on examination of results from a few data sets. The chosen sub-intervals are as follows: [1, 34], [34.5, 67.5] and [68, 100].

A wide range of reasonable initial estimates of $f_1$ and $f_2$ yields good final estimates. For all data sets considered we can always find reasonable initial estimates by eye. For most data sets the proposed automatic methods work. However, for a few data sets the automatic procedures produce obviously bad initial estimates that do not allow the method to recover.

To obtain an initial estimate of $\sigma^2$ we first use the initial function estimates and the two temporary groups of observations to obtain the sample variance of the residuals of each group separately adjusting for the correct degrees of freedom. We then set the initial estimate of $\sigma^{2}$ equal to the pooled variance.

\subsection{Choice of the smoothing parameters, the \texorpdfstring{$\lambda_j$'s}{lambdas}}\label{choiceLambda}

We find the optimal $\lambda_j$'s iteratively starting with initial values $\lambda_j^{(0)}$, $j=1,\ldots,J$. The choice of the $\lambda_j^{(0)}$'s is important to obtain good final estimates of the $f_j$'s. We found out that very small $\lambda_j^{(0)}$'s do not lead to good final estimates as more points tend to be initially misclassified. Recall that for the Bayesian approach we restrict the smoothing parameters of the initial function estimates to be greater than 10. So, for the Bayesian approach, in each data set we set the $\lambda_j^{(0)}$'s to the values used to obtain the initial function estimates. For the penalized log-likelihood approach, we use one value of $\lambda_j^{(0)}$ for all data sets: we set $\lambda_j^{(0)}$ to a value that worked well when tested on a couple of simulated data sets.

\noindent We update the $\lambda_j$'s as follows.

\begin{enumerate}
\item With $\lambda_j=\lambda_j^{(i)}$, $j=1,\ldots,J$, use the EM algorithm of Section \ref{EMsection} to find the $\hat{p}_{ij}$'s, $\hat{\sigma}^2$ and the $\hat{f}_j$'s.
    \item Discard the $\hat{f}_j$'s from Step 1.
\item Treat $\hat{\sigma}^2$ and the $\hat{p}_{ij}$'s as fixed and thus $\matr{W}_j$ as in (\ref{Wjdef}) as fixed. For each $\lambda$ on a grid $\mathcal{G}$ and each $j=1,\ldots,J$, calculate
$$\hat{f}^{\lambda}_j(\vect{x}) = \matr{H}_j(\lambda) \vect{y},$$

\noindent where for the Bayesian approach $\matr{H}_j(\lambda) = \matr{A}(\lambda) ( \matr{A}(\lambda) + \matr{W}_j^{-1})^{-1},$ and for the penalized log-likelihood approach $\matr{H}_j(\lambda) = \matr{B}(\matr{B}^T\matr{W}_j\matr{B} + 2\lambda\matr{R})^{-1}\matr{B}^T\matr{W}_j$.
\item For each $j=1,\ldots,J$, set $\lambda_j^{(i+1)}$ as the value in the grid $\mathcal{G}$ that maximizes the following generalized cross-validation criterion:
\begin{equation*}
GCV_j(\lambda) = \frac{1}{n}\sum_{i=1}^n \hat{p}_{ij} \left ( \frac{ y_i - \hat{f}^{\lambda}_j(x_i)}{1-(H_{j\lambda})_{ii}}  \right)^2,
\end{equation*}
\noindent where $(H_{j\lambda})_{ii}$ is the $i$th entry of the diagonal of $\matr{H}_j(\lambda)$.
\item Repeat 1-4 with $\lambda_j = \lambda_j^{(i+1)}$, $j=1,\ldots,J$, till convergence.

\end{enumerate}

\noindent We use the final values of the $\lambda_j$'s to obtain all of the parameter estimates from the EM algorithm as in Step 1.

\subsection{Results}

Figures \ref{S1Curves}, \ref{S2Curves} and \ref{S3Curves} show the fitted values $\hat{f}_1(\vect{x})$ and $\hat{f}_2(\vect{x})$ (dashed lines) for a simulated data set from each of simulation studies 1, 2 and 3, respectively.

\begin{figure}
        \centering

        \begin{subfigure}[b]{0.45\textwidth}
                \centering
                \includegraphics[width=\textwidth]{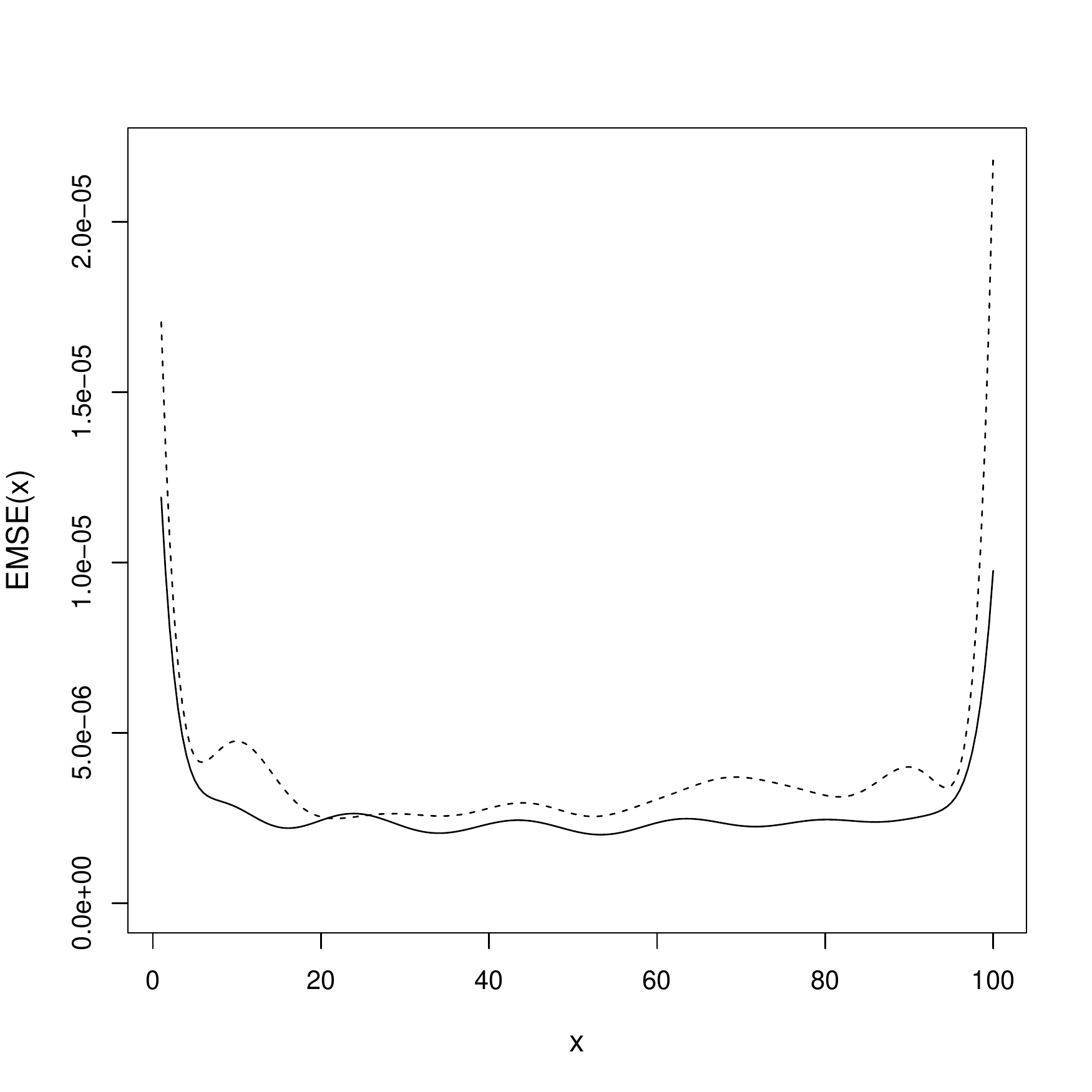}
                \caption{$f_1$: Bayesian}
        \end{subfigure}
        \begin{subfigure}[b]{0.45\textwidth}
                \centering
                \includegraphics[width=\textwidth]{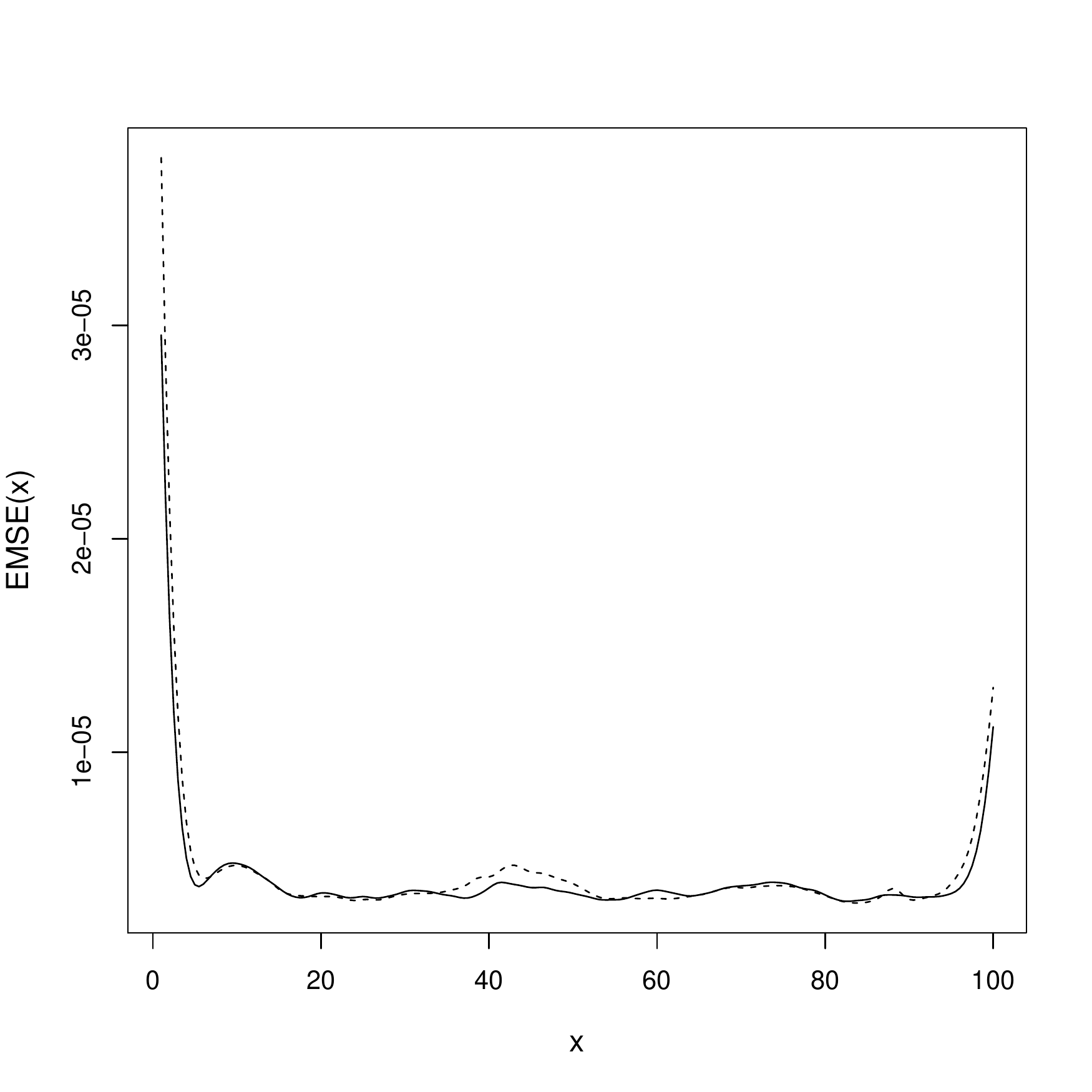}
                \caption{$f_1$: penalized log-likelihood}
        \end{subfigure}
        ~
        \begin{subfigure}[b]{0.45\textwidth}
                \centering
                \includegraphics[width=\textwidth]{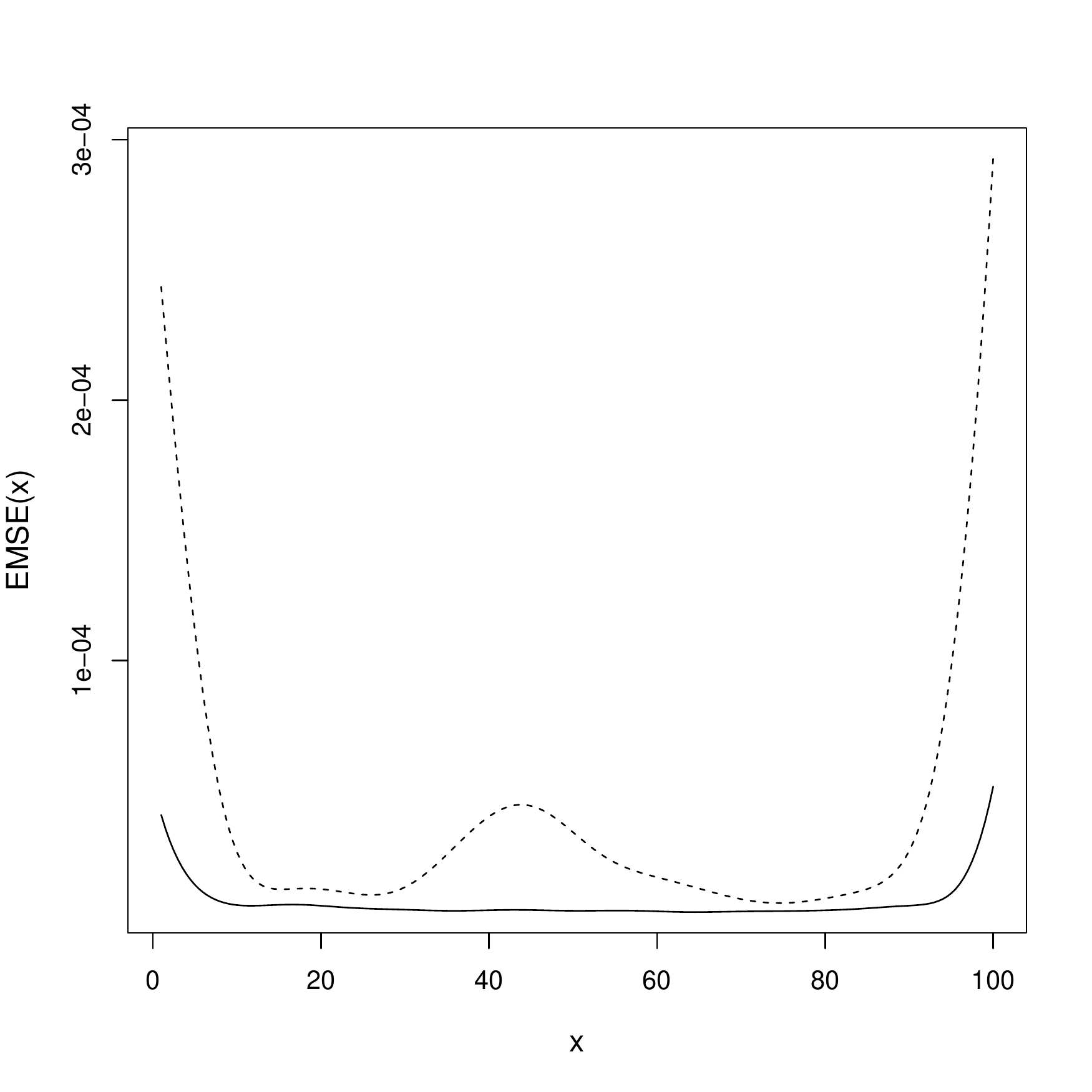}
                \caption{$f_2$: Bayesian}
        \end{subfigure}
        \begin{subfigure}[b]{0.45\textwidth}
                \centering
                \includegraphics[width=\textwidth]{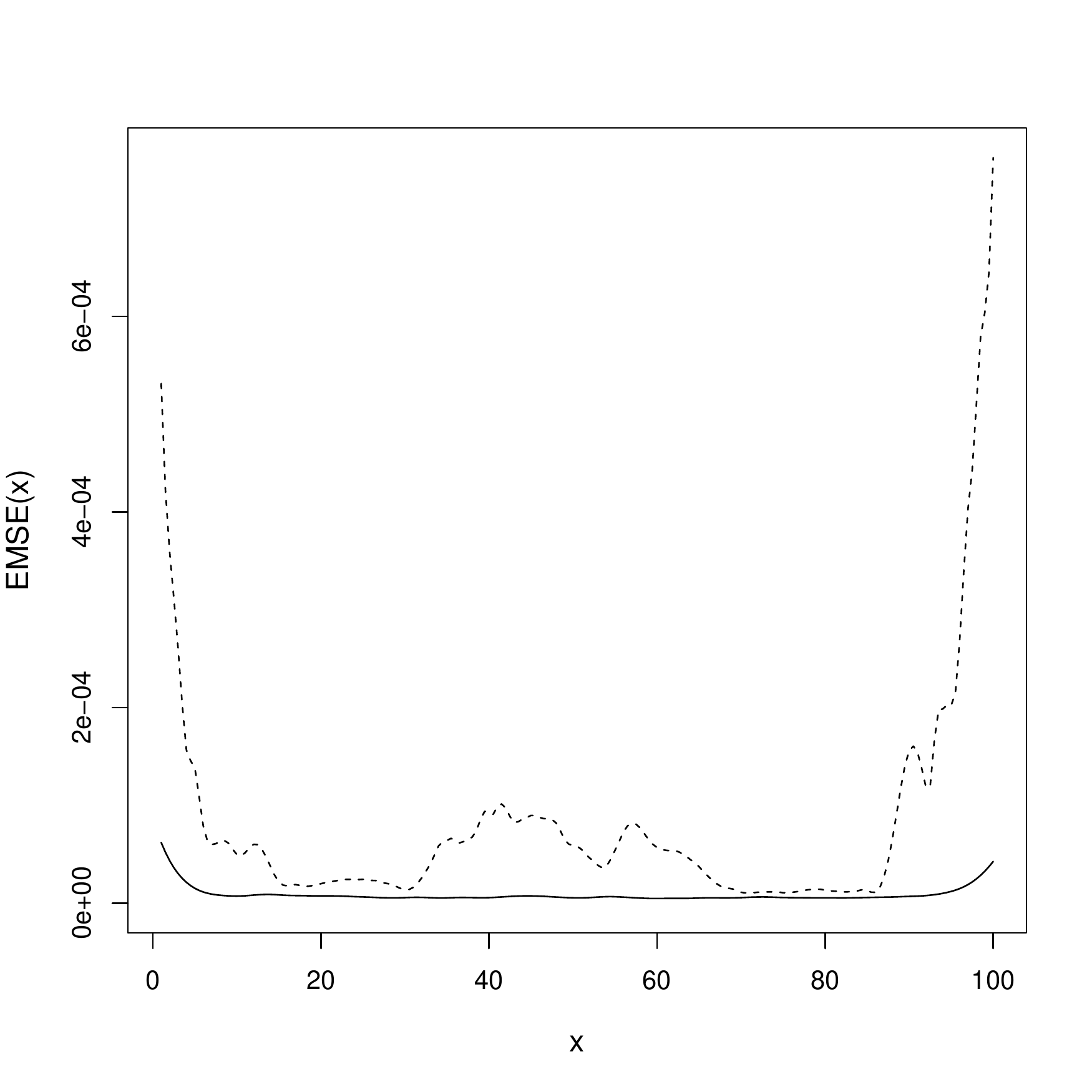}
                \caption{$f_2$: penalized log-likelihood}
        \end{subfigure}
        ~ 
\caption{Simulation 1 (\textit{iid} $z_i$'s). EMSE of both initial (dashed lines) and final (solid lines) estimates of $f_1$ and $f_2$ using the Bayesian and the penalized log-likelihood approaches. Plots (a) and (b) show the results for $f_1$ and (c) and (d) the results for $f_2$.}
\label{S1MSE}
\end{figure}

\begin{figure}
        \centering

        \begin{subfigure}[b]{0.45\textwidth}
                \centering
                \includegraphics[width=\textwidth]{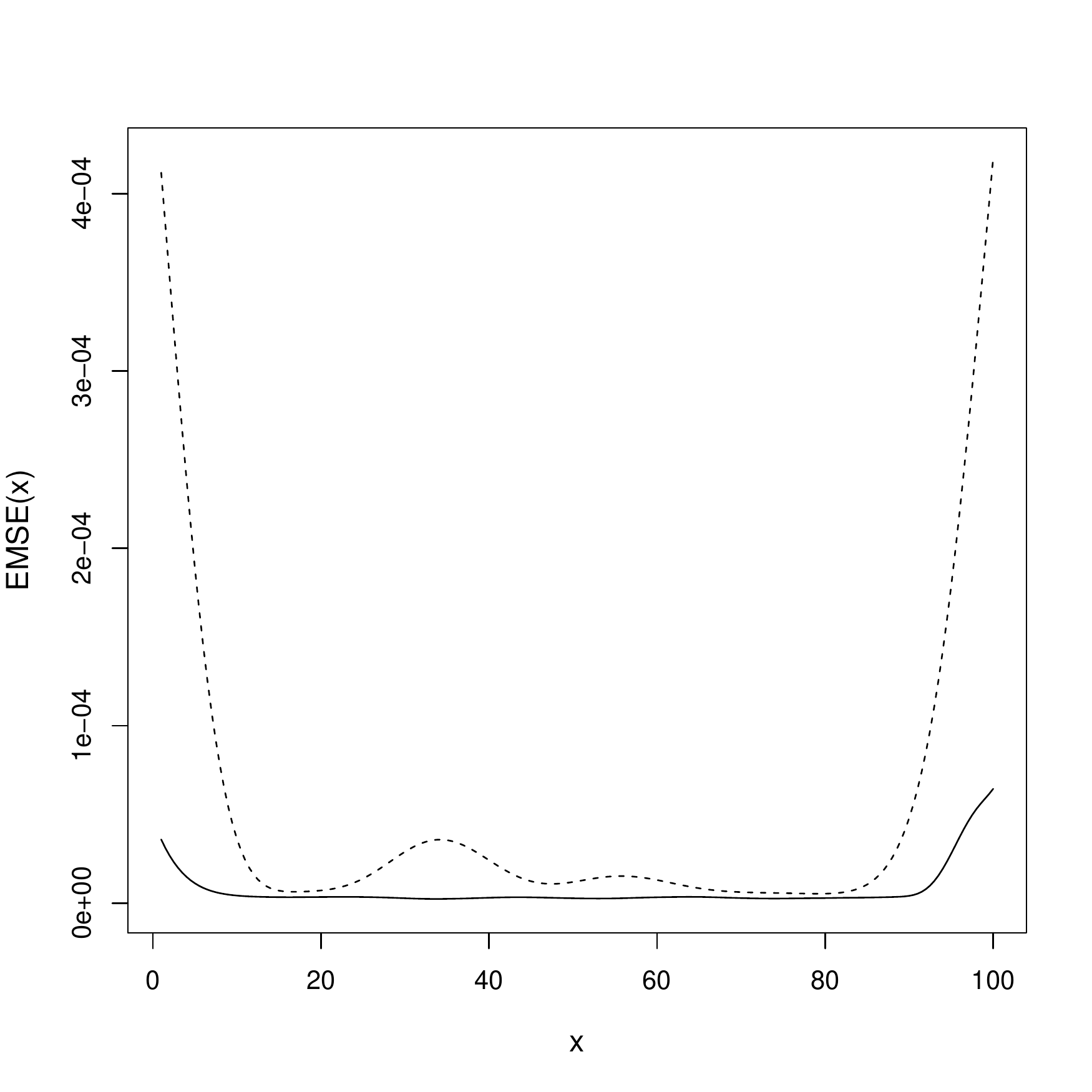}
                \caption{$f_1$: Bayesian}
        \end{subfigure}
        \begin{subfigure}[b]{0.45\textwidth}
                \centering
                \includegraphics[width=\textwidth]{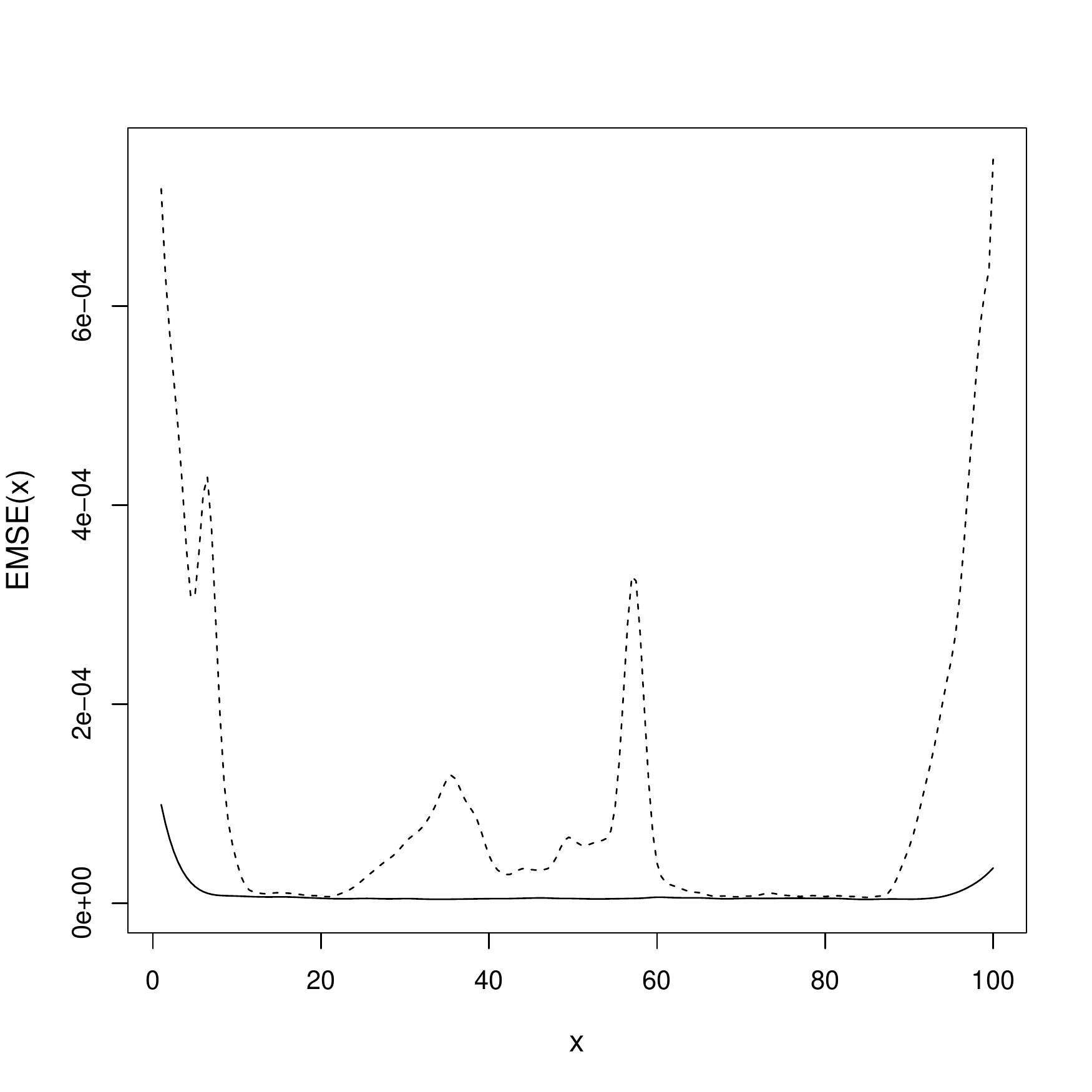}
                \caption{$f_1$: penalized log-likelihood}
        \end{subfigure}
        ~
        \begin{subfigure}[b]{0.45\textwidth}
                \centering
                \includegraphics[width=\textwidth]{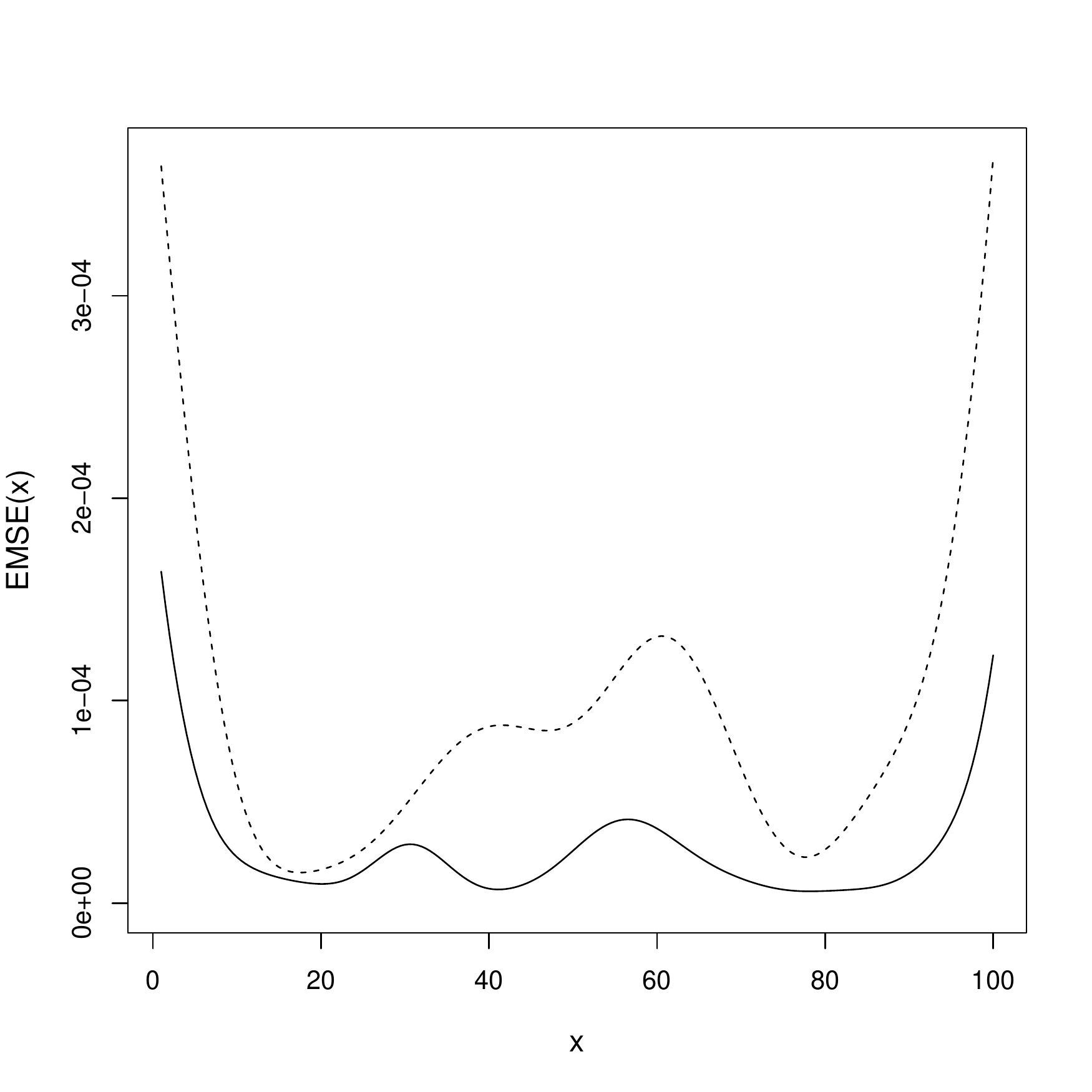}
                \caption{$f_2$: Bayesian}
        \end{subfigure}
        \begin{subfigure}[b]{0.45\textwidth}
                \centering
                \includegraphics[width=\textwidth]{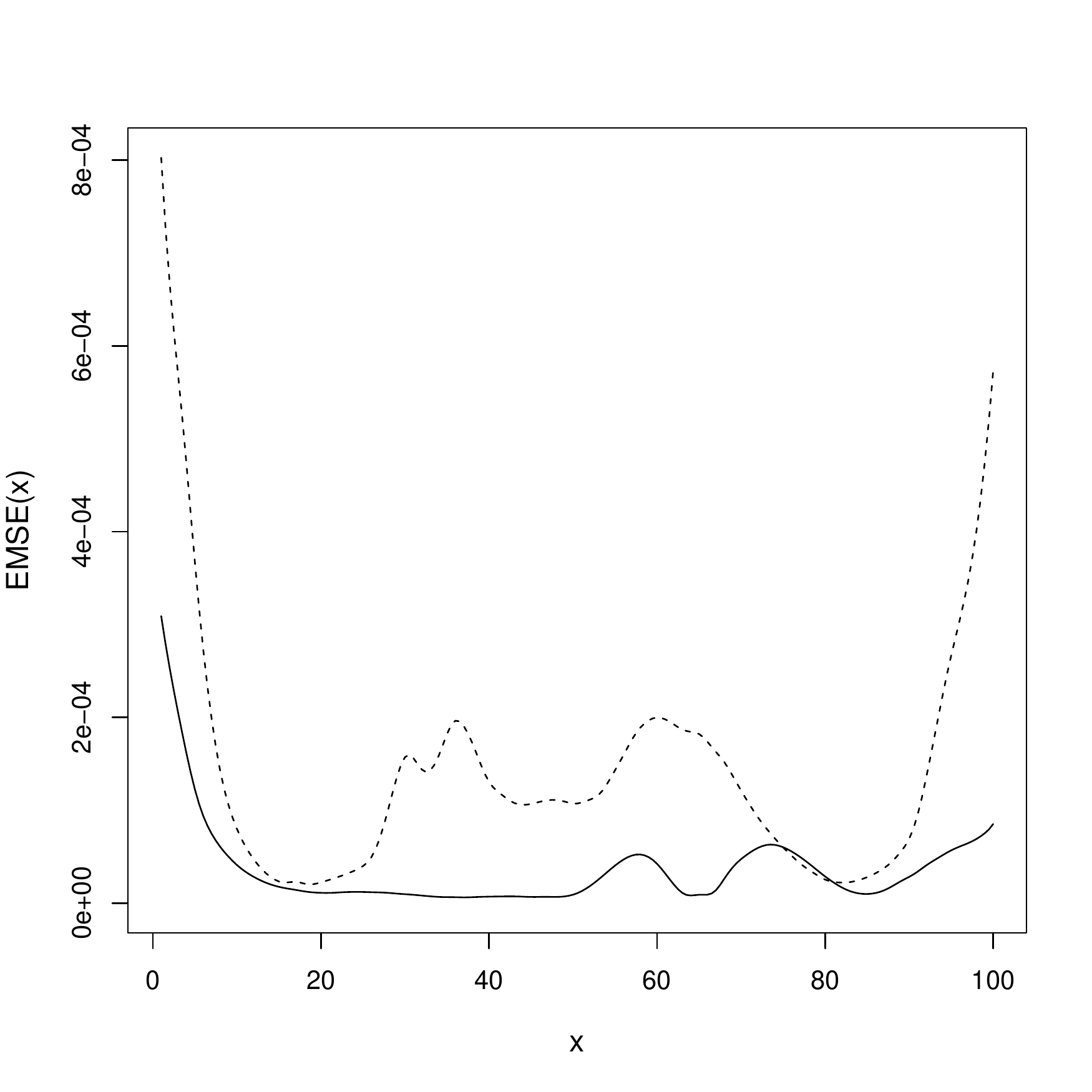}
                \caption{$f_2$: penalized log-likelihood}
        \end{subfigure}
        ~ 
\caption{Simulation 3 (Markov $z_i$'s with $a_{12}=0.1$ and $a_{21}=0.2$). EMSE of both initial (dashed lines) and final (solid lines) estimates of $f_1$ and $f_2$ as in Figure \ref{S1MSE}.}
\label{S3MSE}
\end{figure}

We assess the quality of the estimated functions via the pointwise empirical mean squared error (EMSE). The empirical mean squared error of $\hat{f}_j$ at a given point $x_i$ is calculated as follows:
$$EMSE_j(x_i)= \frac{1}{S}\sum_{s=1}^S \big[ \hat{f}_j^s(x_i)- f_j(x_i)\big]^2,$$

\noindent where $\hat{f}_j^s$ is the estimate of $f_j$ in the $s$th simulated data set and $S$ is the total number of simulated data sets, in this case $S=300$. We calculate the EMSE for both initial and final estimates of the $f_j$'s. In all three simulation studies we observe the presence of edge effects, that is, the $EMSE_j$ values are higher at the edges than at the middle of the evaluation interval. We also observe that in all simulation studies the final estimates produce smaller values of $EMSE_j$ than the initial estimates, which indicates that the proposed methodology improves the initial naive estimates. Figures \ref{S1MSE} and \ref{S3MSE} present the pointwise EMSE of both initial and final estimates of $f_1$ and $f_2$ using the Bayesian and penalized log-likelihood approaches for simulation studies 1 and 3, respectively. The plots for Simulation 2 are omitted because they are very similar to the ones obtained for Simulation 1.

\begin{figure}
        \centering
        \begin{subfigure}[b]{0.5\textwidth}
                \centering
                \includegraphics[width=\textwidth]{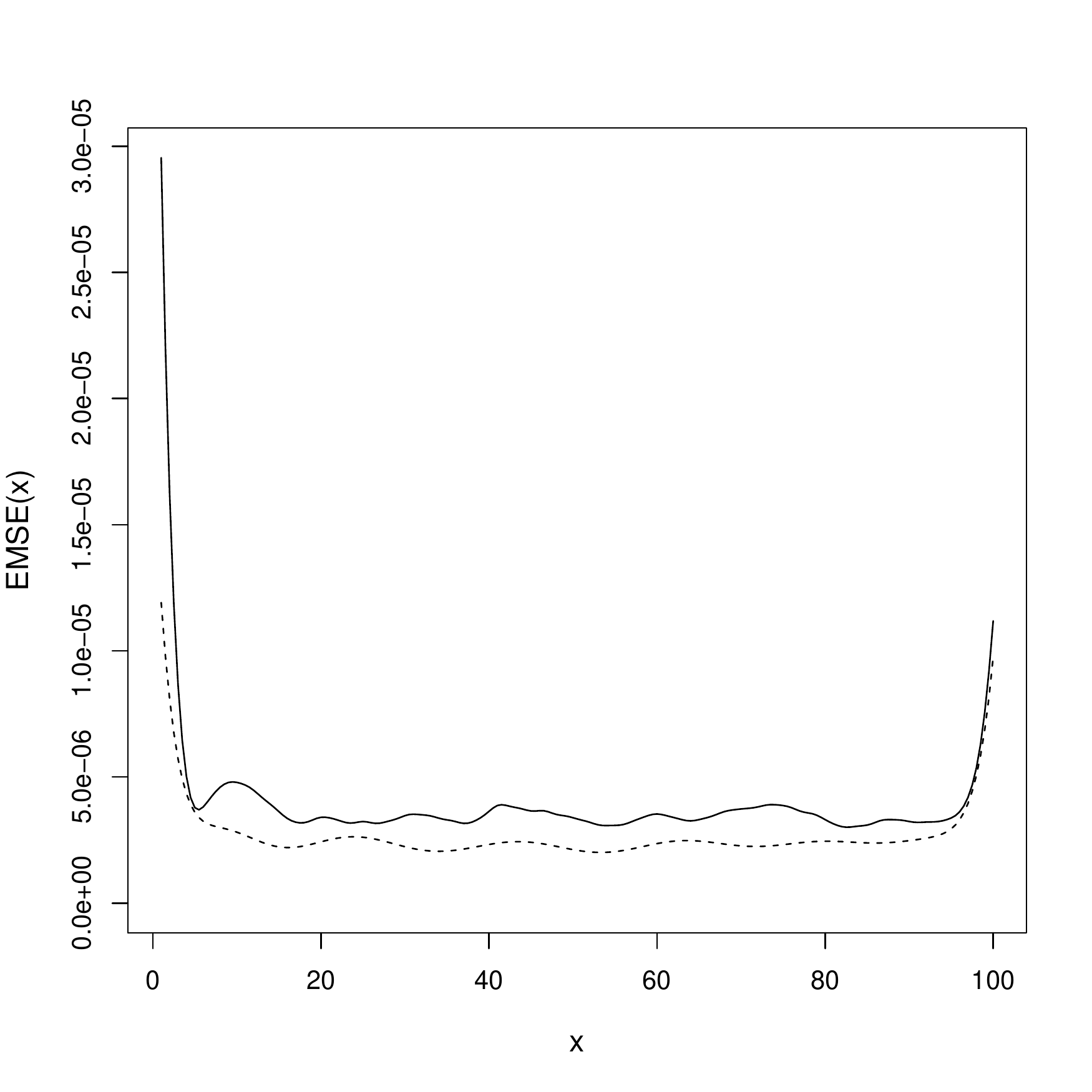}
                \caption{EMSE $\hat{f}_1$}
        \end{subfigure}%
        \begin{subfigure}[b]{0.5\textwidth}
                \centering
                \includegraphics[width=\textwidth]{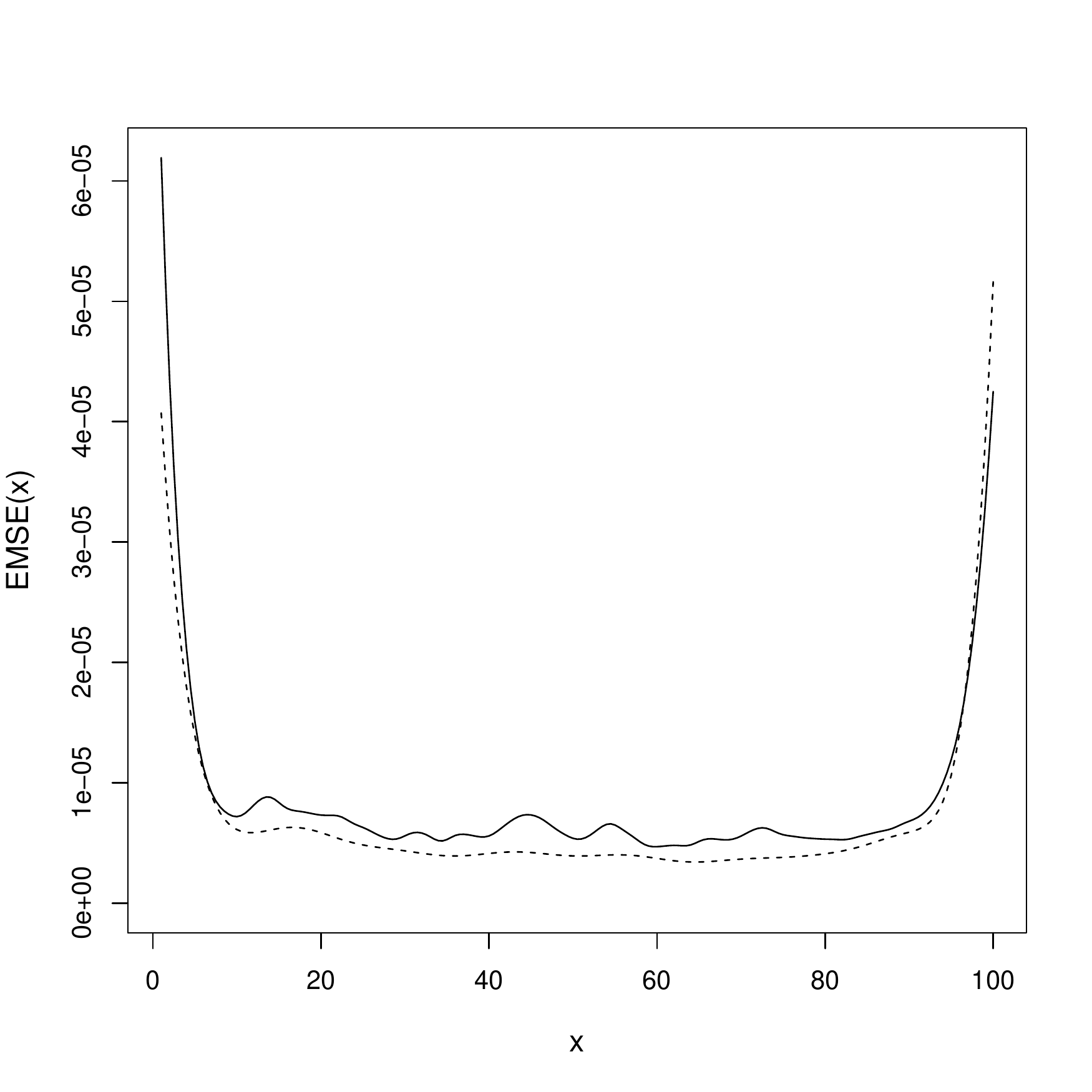}
                \caption{EMSE $\hat{f}_2$}
        \end{subfigure}
\caption{Simulation 1 (\textit{iid} $z_i$'s). EMSE of the final estimates of (a) $f_1$ and (b) $f_2$ using the Bayesian (dashed lines) and the penalized log-likelihood (solid lines) approaches.}
\label{S1MSEBayesVSPL}
\end{figure}

\begin{figure}
        \centering
        \begin{subfigure}[b]{0.5\textwidth}
                \centering
                \includegraphics[width=\textwidth]{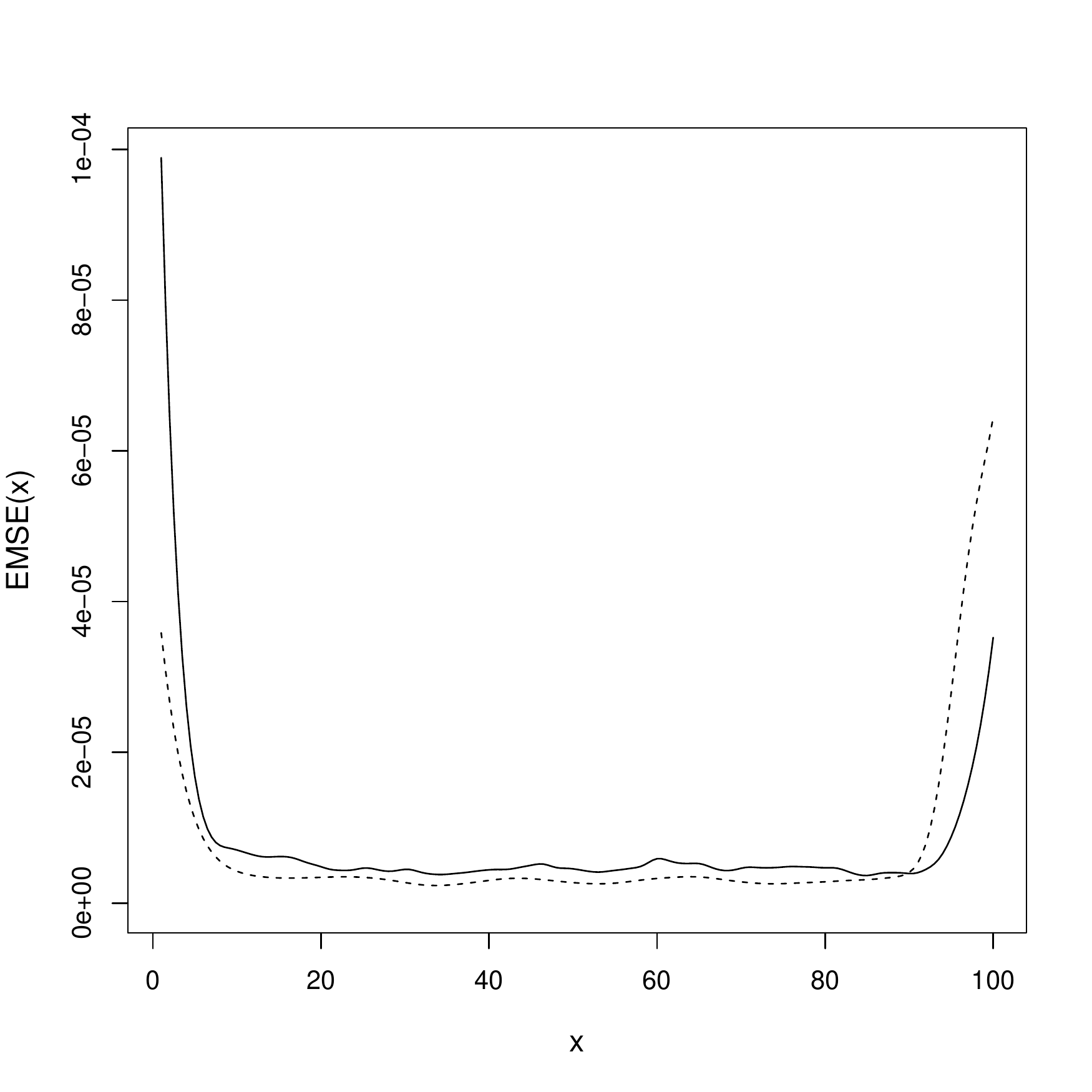}
                \caption{EMSE $\hat{f}_1$}
        \end{subfigure}%
        \begin{subfigure}[b]{0.5\textwidth}
                \centering
                \includegraphics[width=\textwidth]{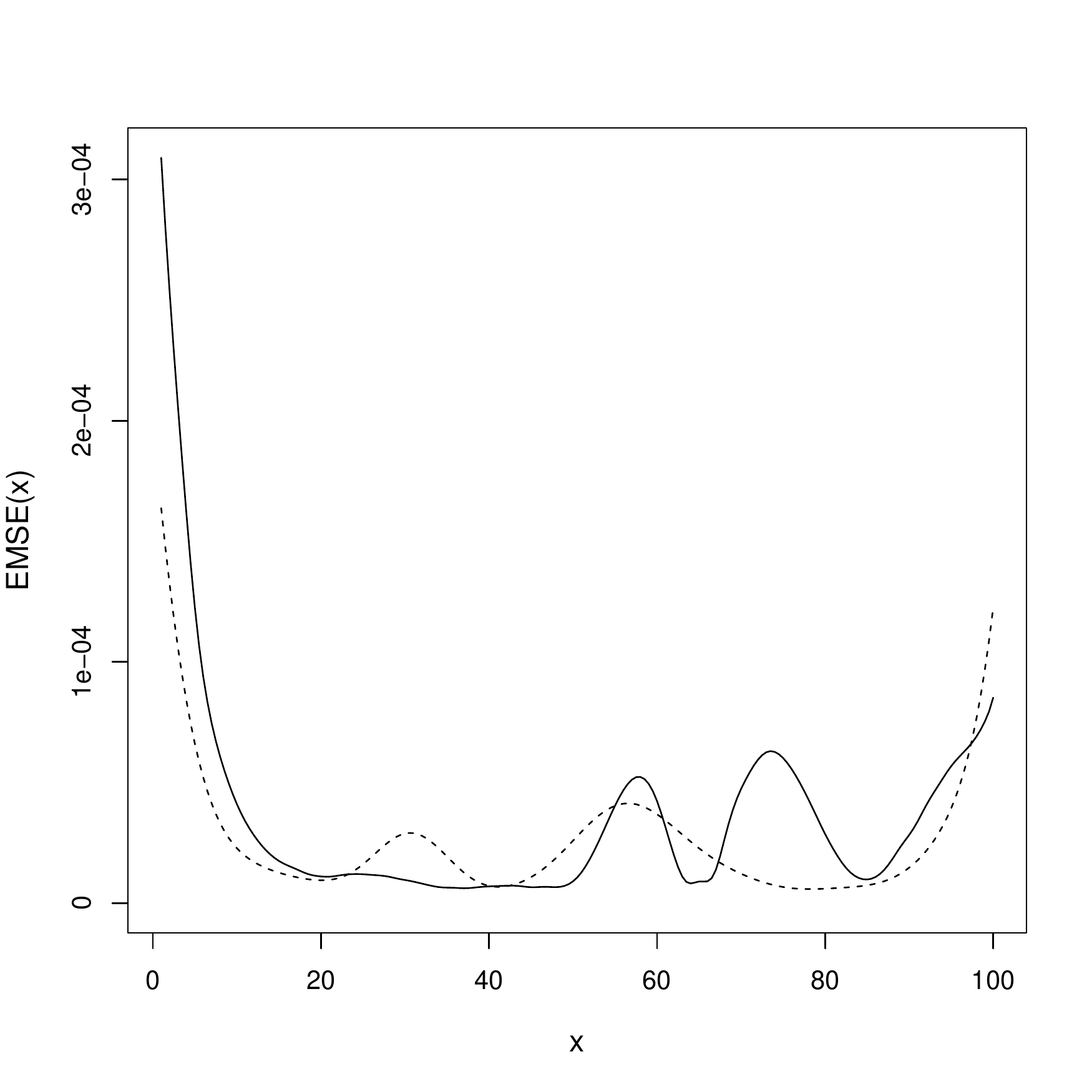}
                \caption{EMSE $\hat{f}_2$}
        \end{subfigure}
\caption{Simulation 3 (Markov $z_i$'s with $a_{12}=0.1$ and $a_{21}=0.2$). EMSE of the final estimates of (a) $f_1$ and (b) $f_2$ as in Figure \ref{S1MSEBayesVSPL}.}
\label{S3MSEBayesVSPL}
\end{figure}

When we compare the Bayesian and the penalized log-likelihood approaches we observe that for Simulations 1 and 2 the Bayesian approach produces slightly smaller values of $EMSE_j$ for the final function estimates than the penalized log-likelihood approach. In Simulation 3 the Bayesian approach produces slightly smaller values of EMSE for $\hat{f}_1$ than the penalized log-likelihood approach except for the right edge. In Simulation 3, there is no clear winner for estimating $f_2$. In addition, both approaches produce values of EMSE for $\hat{f}_2$ that are larger than the values obtained in Simulations 1 and 2. See Figures \ref{S1MSEBayesVSPL} and \ref{S3MSEBayesVSPL}. The plots from Simulation 2 are again omitted as they are very similar to the ones from Simulation 1.

To further study the quality of our method, we look at possible cases of misclassification. We consider values of $\hat{p}(z_i=z|y_i) > 0.2$ when the true $z_i \neq z$. Table \ref{tb:sim:misclass1} presents the number of simulated data sets that have values of $\hat{p}(z_i=1|y_i) > 0.2$ when $z_i=2$. Table \ref{tb:sim:misclass2} shows the number of simulated data sets with values of $\hat{p}(z_i=2|y_i) > 0.2$ when $z_i=1$. For Simulations 1 and 2 all misclassifications occur at the edges of the evaluation interval, indicating that the proposed method sometimes has problems at the edges. The same is true in Simulation 3 for the penalized log-likelihood approach. In Simulation 3, the Bayesian approach leads to problems not just at the edges: two data sets have large values of $\hat{p}(z_i=2|y_i) > 0.2$ when $z_i=1$ at the middle of the evaluation interval when $f_1$ and $f_2$ are closer.

Table \ref{tb:sigma2:summary} presents the mean and standard deviation of all 300 estimates of $\sigma^2$ for each simulation study considering both the Bayesian and the penalized log-likelihood approaches. All estimates are obtained adjusting the degrees of freedom to account for the estimation of the $f_j$'s. We can observe that the means for the Bayesian approach are closer to the true value of $\sigma^2$ ($5\times10^{-5}$) than the means obtained using the penalized log-likelihood approach. The medians (not included in the table) for the Bayesian approach are also closer to $5\times10^{-5}$  than the medians for the penalized log-likelihood approach.

Table \ref{tb:zparam:summary} contains the mean and the standard deviation of the estimates of the parameters of the $z_i$'s for each simulation study considering both the Bayesian and the penalized log-likelihood approaches. Note that the standard deviations of the estimates are close to the values of the means of the proposed standard errors (s.e.'s), as desired. Table \ref{tb:zparam:summary} also shows the empirical coverage percentages of both a 90\% and a 95\% confidence interval. We consider confidence intervals of the form
$$\mbox{mean of the parameter estimates} \; \pm z_{\alpha/2}\times \mbox{proposed s.e.}, $$

\noindent where $z_{\alpha/2}$ is the $\alpha/2$ quantile of a standard normal distribution with $\alpha=0.1$ and 0.05.  The empirical coverage percentages for Simulations 1 and 2 are very close to the true level of the corresponding confidence interval. In Simulation 3 this is not case.
In particular, some of the 90\% confidence intervals for $a_{21}$ are based on estimates of $a_{21}$ that are so poor that even the 95\% confidence intervals do not contain the true value of $a_{21}$. In this case the 95\% confidence intervals have roughly the same empirical coverage as the 90\% confidence intervals.

\section{The motorcycle data revisited}\label{sec:application}

\subsection{Data set background}

The so-called motorcycle data set (Figure \ref{motorcycle}) is a well-known and widely used data set, especially in the fields of nonparametric regression and machine learning. It consists of $n=133$ measurements of head acceleration (in $g$) taken through time (in milliseconds) after impact in simulated motorcycle accidents. A table containing the raw data can be found in \citet{hardle1990applied} and the data are also available in the software R.

The data were collected by \citet{schmidt1981} and became very popular after appearing in \citet{silverman1985some}.
Since then many different methodologies have been applied to the motorcycle data. A Google search shows that this data set appears in more than one hundred articles and book chapters.

The motorcycle data set seems to be quite popular as an example among researchers in areas involving choice of smoothing parameter  (e.g., \citealp{hardle1995fast} and \citealp{wood2011fast}), heterogeneity of the variance  (e.g., \citealp{silverman1985some}) and estimation of a regression function (or its first derivative) with jump discontinuities  (e.g., \citealp{gijbels2004}). It is important to say that all these analyses of the motorcycle data have something in common: they all treat the 133 measurements as coming from one simulated accident. However, as we discussed in the introduction, a close examination of the data suggests the measurements are from $J>1$ accidents.

More recently the motorcycle data set has become a benchmark data set for machine learning techniques involving mixtures of Gaussian processes (e.g., \citealp{yang2011}, \citealp{schiegg2012} and \citealp{lazaro2012}). These techniques still treat the data as coming from one simulated accident and use the different Gaussian processes only as a mechanism to account for the heterogeneity of the variance due to different phases in the data. Therefore, they do not consider the possibility that the different processes in the mixture can actually correspond to multiple runs of accidents. As an example consider the work done by \citet{schiegg2012}. These authors present a machine learning technique called Markov logic mixture of Gaussian processes. They apply their proposed methodology to the motorcycle data in order to fit a single function that takes into account three phases in the data, which they call riding, impact and hitting the ground. Indeed, when, for comparison with their method, \citeauthor{schiegg2012} fit a mixture of three Gaussian processes (or experts) as in \citet{tresp} they write ``the experts of the mixture of Gaussian processes have no specific meaning."

In this section we fit our proposed methodology to the motorcycle data set treating the measurements as coming from $J>1$ functions (one for each simulated accident) with hidden (unknown) function labels. We choose $J$ using an ad hoc Akaike's information criterion (AIC). Unlike the machine learning literature, we provide standard errors for the parameters governing the latent switching process.

\subsection{Data analysis}

We fit the proposed switching nonparametric regression model to the motorcycle data assuming that the hidden states, which correspond to the unknown accident run labels, are \textit{iid}. We estimate the model parameters using both the Bayesian and the penalized log-likelihood approaches. The smoothing parameters, the $\lambda_j$'s, are selected by generalized cross-validation as in Section \ref{choiceLambda}. For the Bayesian approach we set the covariance parameter $U_j$ in (\ref{cov}) to be fixed at
$$U_j = U = \frac{\sum_{i=1}^n(y_i - \bar{y})^2}{n-1}  - \frac{\sum_{i=1}^n(y_i - \hat{f}(x_i))^2}{n-\mbox{trace}(\matr{H})},$$

\noindent where $\hat{f}$ is a regular smoothing spline fit to the data and $\matr{H}$ is the corresponding  hat (or smoothing) matrix. So, $U_j$ is ``known" and $\lambda_j = s_j$.

The data set contains 39 time points with multiple acceleration measurements, which our method does not currently allow. So, for a direct application of our methodology to the data, we jitter each of those time points by adding a very small random noise.

We fit the model considering $J=2,\ldots,6$ functions. We choose $J$ to minimize the ad hoc AIC
\begin{equation}
 -2\log p(\vect{y}|\hat{\theta}) + 2\Big( \sum_{j=1}^J \mbox{trace} (\matr{H}_j ) + \mbox{number of estimated variances} + J-1 \Big), \nonumber
\end{equation}

\noindent where $\matr{H}_j$ is the hat matrix corresponding to $\hat{f}_j(\vect{x})$.

We apply the proposed model considering both equal and different regression error variances. However, the choice of $J$ is not so obvious when we use a common regression error variance. Therefore, the results presented here are obtained considering different error variances.

\begin{figure}
        \centering
        \begin{subfigure}[b]{0.5\textwidth}
                \centering
                \includegraphics[width=\textwidth]{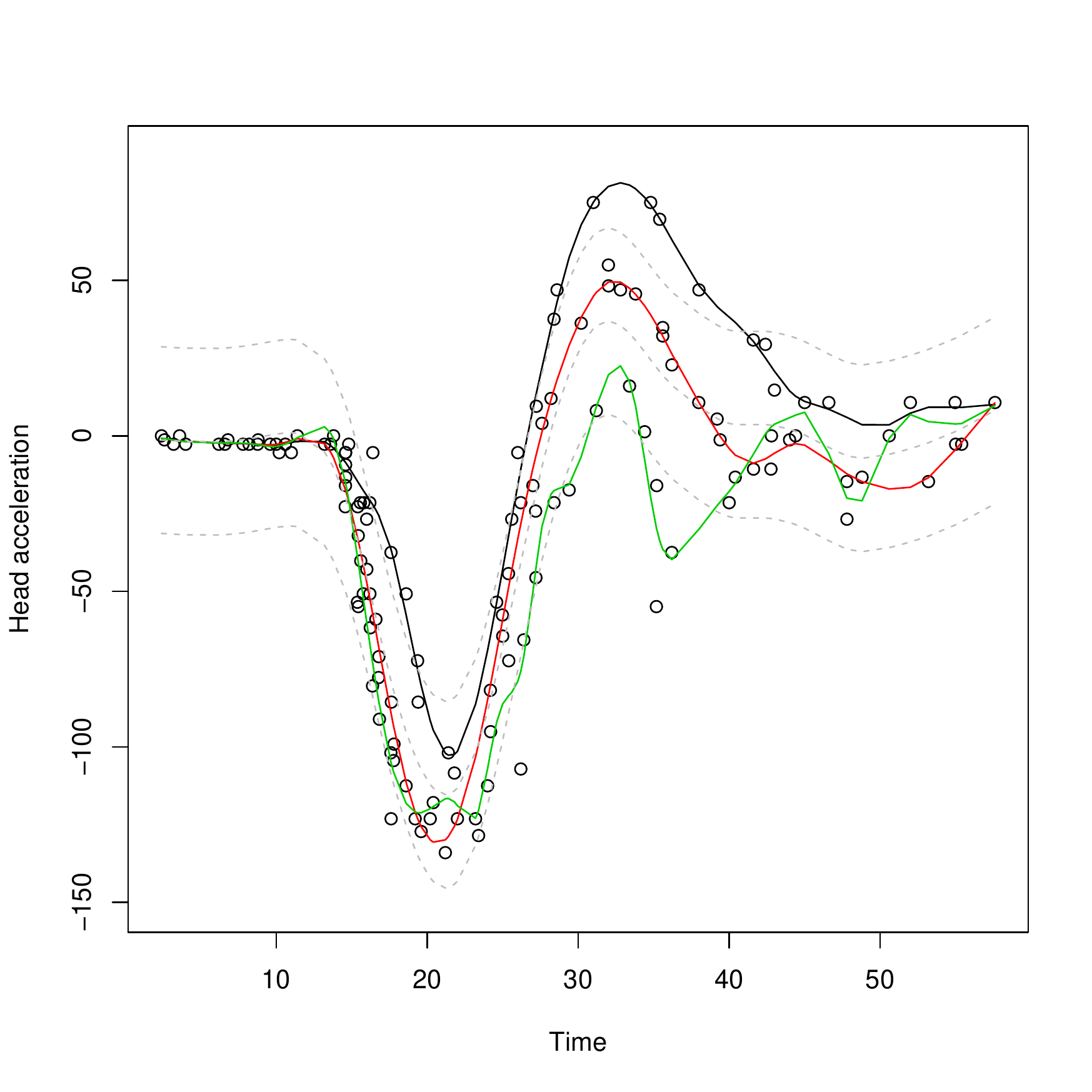}
                \caption{$J=3$}
                \label{3statesPLall}
        \end{subfigure}%
        ~ 
        \begin{subfigure}[b]{0.5\textwidth}
                \centering
                \includegraphics[width=\textwidth]{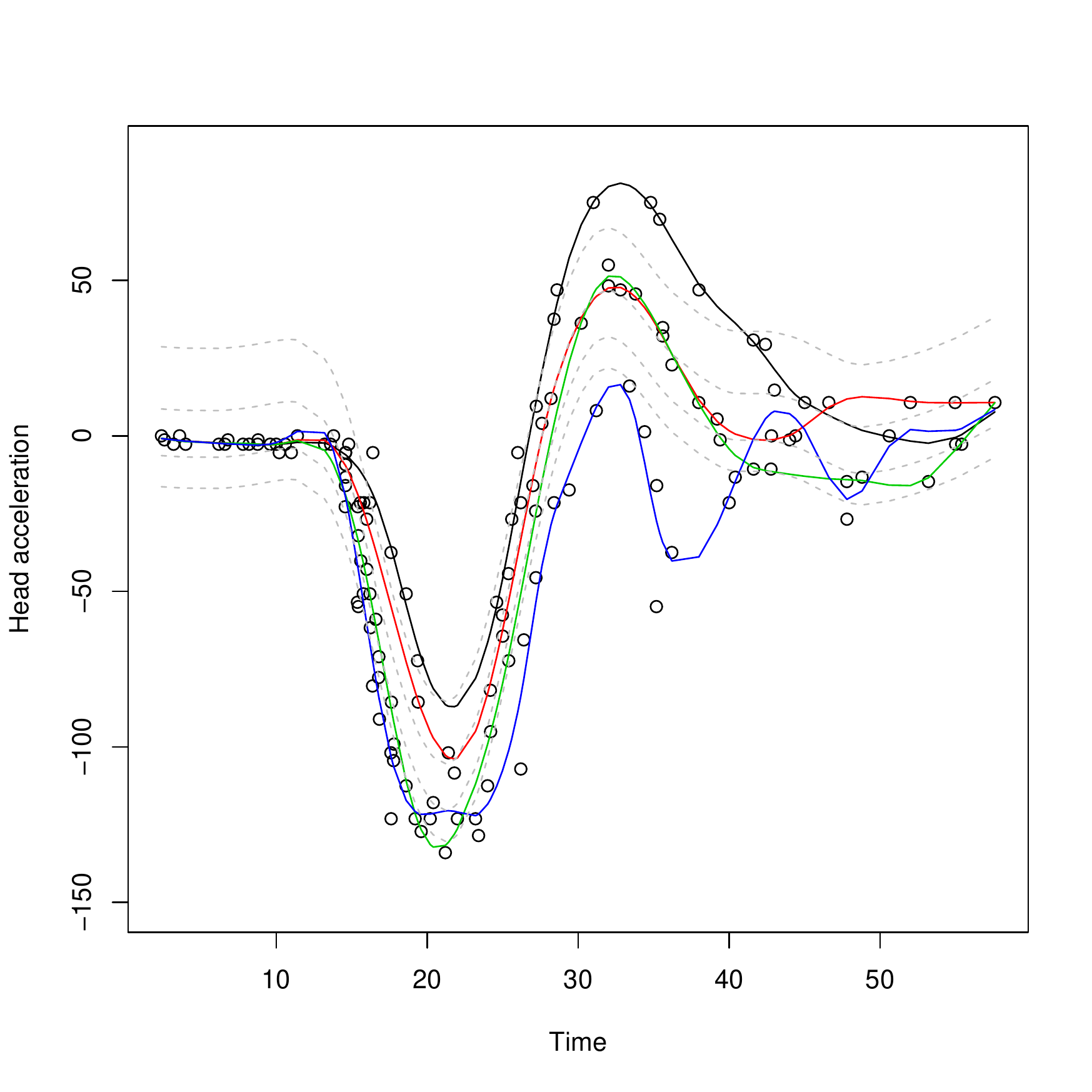}
                \caption{$J=4$}
                \label{4statesPLall}
        \end{subfigure}
\caption{Motorcycle data. Final function estimates (solid lines) obtained using the penalized log-likelihood approach for (a) $J=3$ and (b) $J=4$. The gray dashed lines correspond to the initial function estimates, which are constant shifts of a smoothing spline fit to all the data.}
\label{motorPLCV}
\end{figure}

\begin{figure}
        \centering
              \begin{subfigure}[b]{0.5\textwidth}
                \centering
                \includegraphics[width=\textwidth]{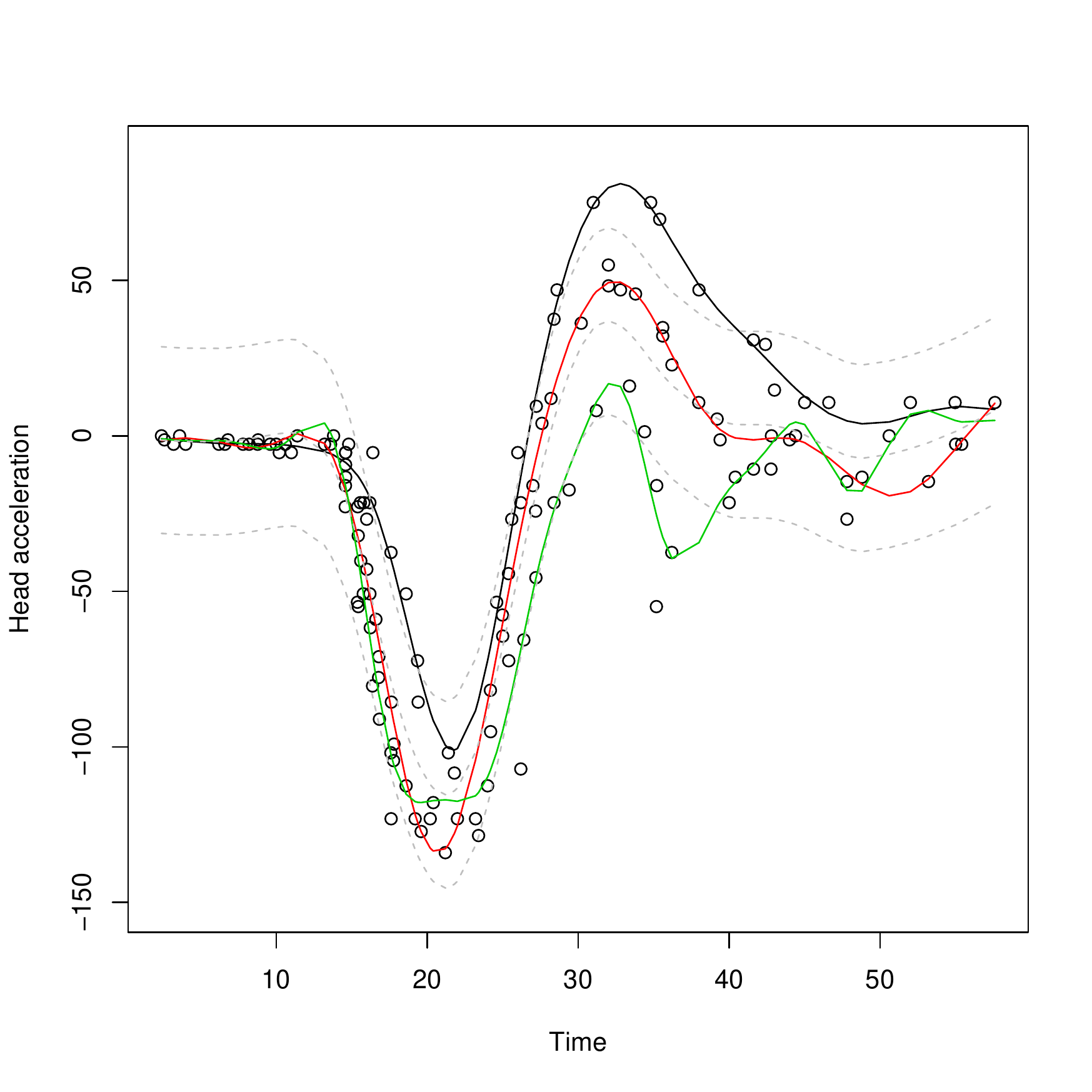}
                \caption{$J=3$}
                \label{3statesBall}
        \end{subfigure}%
        ~ 
        \begin{subfigure}[b]{0.5\textwidth}
                \centering
                \includegraphics[width=\textwidth]{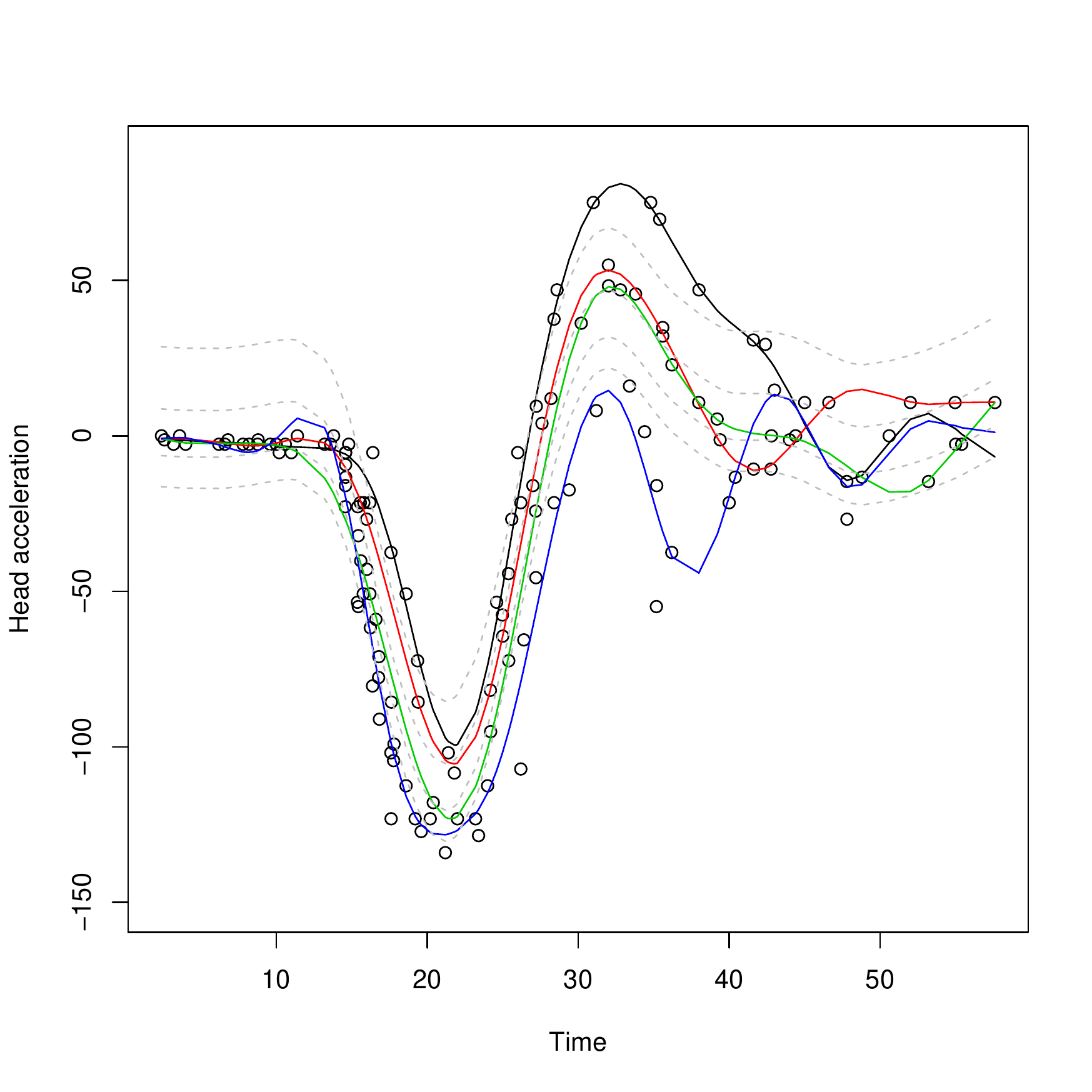}
                \caption{$J=4$}
            \label{4statesBall}
        \end{subfigure}
\caption{Motorcycle data. Final function estimates (solid lines) obtained using the Bayesian approach for (a) $J=3$ and (b) $J=4$. The gray dashed lines correspond to the initial function estimates, which are constant shifts of a smoothing spline fit to all the data.}
\label{motorBayesCV}
\end{figure}

For the Bayesian approach the AIC is minimum for a model with $J=4$. However, for the penalized log-likelihood it is minimum for a model with $J=3$. As both results appear sensible, we present the results for both $J=3$ and $J=4$. Figures \ref{3statesPLall} and \ref{4statesPLall} show the estimated functions obtained using the penalized log-likelihood approach for $J=3$ and $J=4$, respectively, and Figures \ref{3statesBall} and \ref{4statesBall} show the same information using the Bayesian approach.

Table \ref{results3states} presents the estimated model parameters when $J=3$ for both the Bayesian and the penalized log-likelihood approaches. The table also shows the chosen smoothing parameter used in the estimation of each function. We can observe that there is some qualitative agreement between the results from the two approaches. For instance, the green curve has the largest variance. The mixing proportion estimates agree, well within the reported standard errors. Although the values of the $\lambda_j$'s are not comparable between the two approaches, we see that for both methods, the green curve has the smallest value of $\lambda_j$, indicating that the green curve is the least smooth curve.

Table \ref{results4states} is similar to Table \ref{results3states} and presents the results for $J=4$.

\section{Discussion}\label{sec:discussion}
In this paper we proposed a model to analyze data arising from a curve that, over its domain, switches among $J$ states. We called this model a switching nonparametric regression model. Overall our main contributions include the introduction and development of the frequentist approach to the problem, including the calculation of standard errors for the parameter estimators of the latent process and the study of the frequentist properties of the proposed estimates via simulation studies. As an application we analyzed the well-known motorcycle data in an innovative way: treating the data as coming from $J>1$ simulated accident runs with unobserved run labels.
Future work includes the study and development of other criteria to select $J$ and extending our work to a latent process $z$ depending on some covariate(s).

\section*{Acknowledgements}
We would like to thank Prof. Dr. med. Rainer Mattern for all his effort in trying to obtain a copy of \citet{schmidt1981}, the original report containing the motorcycle data. It appears the report is no longer available.

\section*{Supplementary Material}\label{suppA}

\noindent \textbf{Supplement A: The EM and ECM algorithms}\\
\noindent The file \texttt{DeSouzaHeckman-supplementA.pdf} contains the derivation of the EM algorithm used to find the $\hat{\theta}$ that maximizes $\log p(\vect{y}|\theta) + P(f_1,\ldots,f_J,\lambda_1,\ldots,\lambda_J)$. The file also contains a brief description and an example of the ECM algorithm applied in the M-step of the EM algorithm.


\vspace{.5cm}

\noindent \textbf{Supplement B: The \texttt{switchnpreg} package}\\
\noindent Please contact the first author to obtain the R package developed to fit a switching nonparametric regression model. When the required documentation is ready the package \texttt{switchnpreg} will be uploaded on CRAN (\href{http://cran.r-project.org/}{cran.r-project.org}).

\clearpage

\bibliographystyle{imsart-nameyear}
\bibliography{references}

\begin{thebibliography}{24}

\bibitem[\protect\citeauthoryear{Baum et~al.}{1970}]{baum1970}
\begin{barticle}[author]
\bauthor{\bsnm{Baum},~\bfnm{L.~E.}\binits{L.~E.}},
  \bauthor{\bsnm{Petrie},~\bfnm{T.}\binits{T.}},
  \bauthor{\bsnm{Soules},~\bfnm{G.}\binits{G.}} \AND
  \bauthor{\bsnm{Weiss},~\bfnm{N.}\binits{N.}}
(\byear{1970}).
\btitle{A maximization technique occurring in the statistical analysis of
  probabilistic functions of Markov chains}.
\bjournal{The Annals of Mathematical Statistics}
\bvolume{41}
\bpages{164--171}.
\end{barticle}
\endbibitem

\bibitem[\protect\citeauthoryear{Bilmes}{1998}]{bilmes1998}
\begin{barticle}[author]
\bauthor{\bsnm{Bilmes},~\bfnm{J.~A.}\binits{J.~A.}}
(\byear{1998}).
\btitle{A gentle tutorial of the EM algorithm and its application to parameter
  estimation for Gaussian mixture and hidden Markov models}.
\bjournal{International Computer Science Institute}
\bvolume{4}
\bpages{126}.
\end{barticle}
\endbibitem

\bibitem[\protect\citeauthoryear{Capp{\'e}, Moulines and
  Ryd{\'e}n}{2005}]{cappe2005}
\begin{bbook}[author]
\bauthor{\bsnm{Capp{\'e}},~\bfnm{O.}\binits{O.}},
  \bauthor{\bsnm{Moulines},~\bfnm{E.}\binits{E.}} \AND
  \bauthor{\bsnm{Ryd{\'e}n},~\bfnm{T.}\binits{T.}}
(\byear{2005}).
\btitle{Inference in Hidden Markov Models}.
\bpublisher{Springer Verlag}.
\end{bbook}
\endbibitem

\bibitem[\protect\citeauthoryear{Chiou}{2012}]{chiou2012}
\begin{barticle}[author]
\bauthor{\bsnm{Chiou},~\bfnm{J.~M.}\binits{J.~M.}}
(\byear{2012}).
\btitle{Dynamical functional prediction and classification, with application to
  traffic flow prediction}.
\bjournal{The Annals of Applied Statistics}
\bvolume{6}
\bpages{1588--1614}.
\end{barticle}
\endbibitem

\bibitem[\protect\citeauthoryear{Gijbels and Goderniaux}{2004}]{gijbels2004}
\begin{barticle}[author]
\bauthor{\bsnm{Gijbels},~\bfnm{I.}\binits{I.}} \AND
  \bauthor{\bsnm{Goderniaux},~\bfnm{AC}\binits{A.}}
(\byear{2004}).
\btitle{Bootstrap test for change-points in nonparametric regression}.
\bjournal{Journal of Nonparametric Statistics}
\bvolume{16}
\bpages{591--611}.
\end{barticle}
\endbibitem

\bibitem[\protect\citeauthoryear{H{\"a}rdle}{1990}]{hardle1990applied}
\begin{bbook}[author]
\bauthor{\bsnm{H{\"a}rdle},~\bfnm{W.}\binits{W.}}
(\byear{1990}).
\btitle{Applied Nonparametric Regression}.
\bpublisher{Cambridge University Press}.
\end{bbook}
\endbibitem

\bibitem[\protect\citeauthoryear{H{\"a}rdle and Marron}{1995}]{hardle1995fast}
\begin{barticle}[author]
\bauthor{\bsnm{H{\"a}rdle},~\bfnm{W.}\binits{W.}} \AND
  \bauthor{\bsnm{Marron},~\bfnm{J.~S.}\binits{J.~S.}}
(\byear{1995}).
\btitle{Fast and simple scatterplot smoothing}.
\bjournal{Computational Statistics \& Data Analysis}
\bvolume{20}
\bpages{1--17}.
\end{barticle}
\endbibitem

\bibitem[\protect\citeauthoryear{Heckman}{2012}]{heckman2012}
\begin{barticle}[author]
\bauthor{\bsnm{Heckman},~\bfnm{N.}\binits{N.}}
(\byear{2012}).
\btitle{Reproducing Kernel Hilbert Spaces made easy}.
\bjournal{Statistics Surveys}
\bvolume{6}
\bpages{113-141}.
\end{barticle}
\endbibitem

\bibitem[\protect\citeauthoryear{Johnson and Wichern}{2008}]{johnson2008}
\begin{bbook}[author]
\bauthor{\bsnm{Johnson},~\bfnm{R.~A.}\binits{R.~A.}} \AND
  \bauthor{\bsnm{Wichern},~\bfnm{D.~W.}\binits{D.~W.}}
(\byear{2008}).
\btitle{Applied Multivariate Statistical Analysis}.
\bpublisher{6th Ed., Pearson}.
\end{bbook}
\endbibitem

\bibitem[\protect\citeauthoryear{L{\'a}zaro-Gredilla, Van~Vaerenbergh and
  Lawrence}{2012}]{lazaro2012}
\begin{barticle}[author]
\bauthor{\bsnm{L{\'a}zaro-Gredilla},~\bfnm{M.}\binits{M.}},
  \bauthor{\bsnm{Van~Vaerenbergh},~\bfnm{S.}\binits{S.}} \AND
  \bauthor{\bsnm{Lawrence},~\bfnm{N.~D.}\binits{N.~D.}}
(\byear{2012}).
\btitle{Overlapping mixtures of Gaussian processes for the data association
  problem}.
\bjournal{Pattern Recognition}
\bvolume{45}
\bpages{1386--1395}.
\end{barticle}
\endbibitem

\bibitem[\protect\citeauthoryear{Louis}{1982}]{louis1982}
\begin{barticle}[author]
\bauthor{\bsnm{Louis},~\bfnm{T.~A.}\binits{T.~A.}}
(\byear{1982}).
\btitle{Finding the observed information matrix when using the EM algorithm}.
\bjournal{Journal of the Royal Statistical Society Series B}
\bvolume{44}
\bpages{226--233}.
\end{barticle}
\endbibitem

\bibitem[\protect\citeauthoryear{McLachlan and Krishnan}{2008}]{mclachlan}
\begin{bbook}[author]
\bauthor{\bsnm{McLachlan},~\bfnm{G.~J.}\binits{G.~J.}} \AND
  \bauthor{\bsnm{Krishnan},~\bfnm{T.}\binits{T.}}
(\byear{2008}).
\btitle{{The EM Algorithm and Extensions}}.
\bpublisher{2nd Ed., Wiley New York}.
\end{bbook}
\endbibitem

\bibitem[\protect\citeauthoryear{Ou and Martin}{2008}]{ou2008}
\begin{barticle}[author]
\bauthor{\bsnm{Ou},~\bfnm{X.}\binits{X.}} \AND
  \bauthor{\bsnm{Martin},~\bfnm{E.}\binits{E.}}
(\byear{2008}).
\btitle{Batch process modelling with mixtures of Gaussian processes}.
\bjournal{Neural {C}omputing \& {A}pplications}
\bvolume{17}
\bpages{471--479}.
\end{barticle}
\endbibitem

\bibitem[\protect\citeauthoryear{Rabiner}{1989}]{rabiner1989}
\begin{barticle}[author]
\bauthor{\bsnm{Rabiner},~\bfnm{L.~R.}\binits{L.~R.}}
(\byear{1989}).
\btitle{A tutorial on hidden Markov models and selected applications in speech
  recognition}.
\bjournal{Proceedings of the IEEE}
\bvolume{77}
\bpages{257--286}.
\end{barticle}
\endbibitem

\bibitem[\protect\citeauthoryear{Rasmussen and
  Ghahramani}{2002}]{rasmussen2002}
\begin{binproceedings}[author]
\bauthor{\bsnm{Rasmussen},~\bfnm{C.~E.}\binits{C.~E.}} \AND
  \bauthor{\bsnm{Ghahramani},~\bfnm{Z.}\binits{Z.}}
(\byear{2002}).
\btitle{Infinite mixtures of Gaussian process experts}.
In \bbooktitle{Advances in {N}eural {I}nformation {P}rocessing {S}ystems 14:
  {P}roceedings of the 2001 {C}onference}
\bvolume{2}
\bpages{881--888}.
\bpublisher{The MIT Press}.
\end{binproceedings}
\endbibitem

\bibitem[\protect\citeauthoryear{Rasmussen and Williams}{2006}]{rasmussen2006}
\begin{bbook}[author]
\bauthor{\bsnm{Rasmussen},~\bfnm{C.~E.}\binits{C.~E.}} \AND
  \bauthor{\bsnm{Williams},~\bfnm{C.~K.~I.}\binits{C.~K.~I.}}
(\byear{2006}).
\btitle{{Gaussian Processes for Machine Learning}}.
\bpublisher{The MIT Press}.
\end{bbook}
\endbibitem

\bibitem[\protect\citeauthoryear{Schiegg, Neumann and
  Kersting}{2012}]{schiegg2012}
\begin{binproceedings}[author]
\bauthor{\bsnm{Schiegg},~\bfnm{M.}\binits{M.}},
  \bauthor{\bsnm{Neumann},~\bfnm{M.}\binits{M.}} \AND
  \bauthor{\bsnm{Kersting},~\bfnm{K.}\binits{K.}}
(\byear{2012}).
\btitle{Markov logic mixtures of Gaussian processes: towards machines reading
  regression data}.
In \bbooktitle{Proceedings of the 15th International Conference on Artificial
  Intelligence and Statistics}
\bvolume{22}.
\end{binproceedings}
\endbibitem

\bibitem[\protect\citeauthoryear{Schmidt, Mattern and
  Sch{\"u}ler}{1981}]{schmidt1981}
\begin{barticle}[author]
\bauthor{\bsnm{Schmidt},~\bfnm{G.}\binits{G.}},
  \bauthor{\bsnm{Mattern},~\bfnm{R.}\binits{R.}} \AND
  \bauthor{\bsnm{Sch{\"u}ler},~\bfnm{F.}\binits{F.}}
(\byear{1981}).
\btitle{Biomechanical investigation to determine physical and traumatological
  differentiation criteria for the maximum load capacity of head and vertebral
  column with and without protective helmet under the effects of impact}.
\bjournal{EEC Research Program on Biomechanics of Impacts. Final report Phase
  III, Project G5, Institut f{\"u}r Rechtsmedizin, Universit{\"a}t Heidelberg}.
\end{barticle}
\endbibitem

\bibitem[\protect\citeauthoryear{Shi, Murray-Smith and
  Titterington}{2005}]{shi2005}
\begin{barticle}[author]
\bauthor{\bsnm{Shi},~\bfnm{J.~Q.}\binits{J.~Q.}},
  \bauthor{\bsnm{Murray-Smith},~\bfnm{R.}\binits{R.}} \AND
  \bauthor{\bsnm{Titterington},~\bfnm{DM}\binits{D.}}
(\byear{2005}).
\btitle{Hierarchical Gaussian process mixtures for regression}.
\bjournal{Statistics and Computing}
\bvolume{15}
\bpages{31--41}.
\end{barticle}
\endbibitem

\bibitem[\protect\citeauthoryear{Silverman}{1985}]{silverman1985some}
\begin{barticle}[author]
\bauthor{\bsnm{Silverman},~\bfnm{B.~W.}\binits{B.~W.}}
(\byear{1985}).
\btitle{Some aspects of the spline smoothing approach to non-parametric
  regression curve fitting}.
\bjournal{Journal of the Royal Statistical Society Series B}
\bvolume{47}
\bpages{1--52}.
\end{barticle}
\endbibitem

\bibitem[\protect\citeauthoryear{Tresp}{2001}]{tresp}
\begin{binproceedings}[author]
\bauthor{\bsnm{Tresp},~\bfnm{V.}\binits{V.}}
(\byear{2001}).
\btitle{Mixtures of Gaussian processes}.
In \bbooktitle{Advances in {N}eural {I}nformation {P}rocessing {S}ystems 13:
  {P}roceedings of the 2000 {C}onference}
\bpages{654--660}.
\bpublisher{The MIT Press}.
\end{binproceedings}
\endbibitem

\bibitem[\protect\citeauthoryear{Wahba}{1983}]{wahba1983bayesian}
\begin{barticle}[author]
\bauthor{\bsnm{Wahba},~\bfnm{G.}\binits{G.}}
(\byear{1983}).
\btitle{Bayesian ``confidence intervals" for the cross-validated smoothing
  spline}.
\bjournal{Journal of the Royal Statistical Society Series B}
\bvolume{45}
\bpages{133--150}.
\end{barticle}
\endbibitem

\bibitem[\protect\citeauthoryear{Wood}{2011}]{wood2011fast}
\begin{barticle}[author]
\bauthor{\bsnm{Wood},~\bfnm{Simon~N}\binits{S.~N.}}
(\byear{2011}).
\btitle{Fast stable restricted maximum likelihood and marginal likelihood
  estimation of semiparametric generalized linear models}.
\bjournal{Journal of the Royal Statistical Society Series B}
\bvolume{73}
\bpages{3--36}.
\end{barticle}
\endbibitem

\bibitem[\protect\citeauthoryear{Yang and Ma}{2011}]{yang2011}
\begin{binproceedings}[author]
\bauthor{\bsnm{Yang},~\bfnm{Y.}\binits{Y.}} \AND
  \bauthor{\bsnm{Ma},~\bfnm{J.}\binits{J.}}
(\byear{2011}).
\btitle{An efficient EM approach to parameter learning of the mixture of
  gaussian processes}.
In \bbooktitle{Proceedings of the 8th International Conference on Advances in
  Neural Networks}
\bpages{165--174}.
\bpublisher{Springer-Verlag}.
\end{binproceedings}
\endbibitem

\end{thebibliography}

\clearpage


\begin{table}[ht]
\caption{Summary of the simulation studies.}
\centering
\begin{tabular}{c| c| c| c| c} 
\thickhline
\multirow{2}{*}{Sim'n} & \multirow{2}{*}{\minitab[c]{type \\ of $z_i$'s}}  & \multirow{2}{*}{\minitab[c]{$z_i$'s \\ parameters}} & \multirow{2}{*}{$\sigma_j^2$'s} & \multirow{2}*{\minitab[c]{$f_j(\vect{x}) \sim $ \\ $MVN(\vect{0},\matr{A}(\lambda_j))$} } \\
 &  &  &  &  \\
\thickhline
 \multirow{2}{*}{1} & \multirow{2}{*}{\textit{iid}}  & $p_1=0.7$  &  \multirow{8}{*}{\minitab[c]{$\sigma_1^2=\sigma_2^2=$ \\ $5\times 10^{-5}$}} & \multirow{8}{*}{\minitab[c]{$\lambda_1=28$ \\ $\lambda_2=38$}}  \\
  &                     &  $p_2 =0.3$ &     &    \\ \cline{1-3}
 \multirow{3}{*}{2} & \multirow{6}{*}{Markov} 			& $\pi_1=\pi_2=0.5$ 						&  & \\
	 &					 		& $a_{12}=0.3$ &  								 &                            	\\ 
	&								&  $a_{21}=0.4$ &  & \\ \cline{1-1}\cline{3-3}
 \multirow{3}{*}{3} &       				& $\pi_1=\pi_2=0.5$ 						&	  & 	\\									
	 &				 			&	$a_{12}=0.1$  & 	 							 &  \\
&									&	$a_{21}=0.2$ & 	 							 &  \\
\thickhline
\end{tabular}
\label{tb:sim:summary}
\end{table}

\begin{table}[ht]
\caption{Number of simulated data sets with values of $\hat{p}(z_i=1|y_i) > 0.2$ when $z_i=2$.}
\centering
\begin{tabular}{c| c| c| c} 
\thickhline
Approach & Simulation 1  & Simulation 2 & Simulation 3  \\
\thickhline
Bayes                       & $0^*$     & $0^{*}$   & 4 \\
Penalized log-likelihood    & $0^{*}$   & $0^{*}$   & 1 \\
\thickhline
\multicolumn{4}{l}{\footnotesize $^*$ $\hat{p}(z_i=1|y_i)$ was never larger than $0.04$ when the true $z_i=2$.}
\end{tabular}
\label{tb:sim:misclass1}
\end{table}

\begin{table}[ht]
\caption{Number of simulated data sets with values of $\hat{p}(z_i=2|y_i) > 0.2$ when $z_i=1$.}
\centering
\begin{tabular}{c| c| c| c} 
\thickhline
Approach & Simulation 1  & Simulation 2 & Simulation 3  \\
\thickhline
Bayes                    & 3 & 3  & 14 \\
Penalized log-likelihood & 2 & 1  & 13 \\
\thickhline
\end{tabular}
\label{tb:sim:misclass2}
\end{table}

\clearpage

\begin{table}
\caption{Estimates of $\sigma^2$ (true value $= 5\times10^{-5}$).}
\centering
\begin{tabular}{c| c| c}
\thickhline
Simulation & approach & mean$\times 10^{5}$ (SD$^{*}\times 10^{5}$) \\
\thickhline
\multirow{2}{*}{1} & Bayesian & 4.984 (0.491) \\
                     & PL$^{**}$ & 4.912 (0.498) \\
\hline
\multirow{2}*{2} &  Bayesian  & 4.982 (0.556) \\
 & PL  & 4.916 (0.579) \\
\hline
\multirow{2}*{3} & Bayesian  & 4.935 (0.517) \\
        & PL & 4.880 (0.519) \\
\thickhline
\multicolumn{3}{l}{\footnotesize $^*$ SD = standard deviation, $^{**}$ PL = penalized log-likelihood. }
\end{tabular}
\label{tb:sigma2:summary}
\end{table}

\begin{table}
\caption{Estimates of the parameters of the $z$ process.}
\centering
\begin{tabular}{c| c| c| c| c | c| c} 
\thickhline
\multirow{2}{*}{Sim'n}  &\multirow{2}*{\minitab[c]{$z_i$'s\\ parameters}} & \multirow{2}{*}{approach} & \multirow{2}{*}{mean (SD$^{*}$)} & mean & \multicolumn{2}{c}{empirical coverage}   \\ \cline{6-7}
 &  & & & of s.e.'s &  90\%  & 95\%  \\
\thickhline
\multirow{2}{*}{1} & \multirow{2}{*}{$p_1=0.7$} & Bayesian & 0.699 (0.032) & 0.032 &  90.7\% & 95.7\% \\
 &  & PL$^{**}$ &  0.699 (0.032) & 0.032 &  90.7\% & 95.7\% \\
\hline
\multirow{4}{*}{2} &  \multirow{2}{*}{$a_{12}=0.3$} &  Bayesian & 0.300 (0.043) & 0.043 & 90.0\% & 94.3\% \\
  & & PL &   0.300 (0.043) & 0.043 & 90.0\% & 94.3\%     \\  \cline{2-7}
  & \multirow{2}{*}{$a_{21}=0.4$} & Bayesian &  0.399 (0.053) & 0.053 & 90.3\% & 95.7\%  \\
  & & PL & 0.399 (0.053)  & 0.053& 90.3\% & 95.7\%   \\
  \hline
\multirow{4}{*}{3} &  \multirow{2}{*}{$a_{12}=0.1$} & Bayesian& 0.105 (0.025)  & 0.027 & 93.0\% & 97.3\% \\
& & PL & 0.105 (0.024) & 0.027 & 93.3\% & 97.3\% \\ \cline{2-7}
 & \multirow{2}{*}{$a_{21}=0.2$} & Bayesian & 0.208 (0.052) & 0.050 & 91.7\% & 92.7\% \\
 & & PL & 0.208 (0.052) & 0.050 & 91.7\% & 92.7\% \\
\thickhline
\multicolumn{7}{l}{\footnotesize $^*$ SD = standard deviation, $^{**}$ PL = penalized log-likelihood. }
\end{tabular}
\label{tb:zparam:summary}
\end{table}

\clearpage



\begin{table}[ht]
\caption{Results for $J=3$ with corresponding fitted curves in Figures \ref{3statesPLall} and \ref{3statesBall}.}
\vspace{-.4cm}
\begin{center}
\begin{tabular}{c|c|c|c|c}
  \thickhline
Approach & curve & $\hat{\sigma}^2_j$ & $\hat{p}_j$ (s.e.) & $\lambda_j$ \ \\
  \thickhline
 \multirow{3}*{\minitab[c]{Penalized\\ log-likelihood \\ (Fig. \ref{3statesPLall})}} &   black  & 43.054 & 0.269  (0.047) & 0.893 \\
   & red   & 14.227 & 0.395 (0.053) & 0.890 \\
   & green & 171.048   & 0.337  (0.053)  & 0.167 \\
   \hline
 \multirow{3}*{\minitab[c]{Bayesian \\ (Fig. \ref{3statesBall})}} &   black     & 50.134    & 0.272 (0.047) & 3.901 \\
                                                                        & red   & 8.593     & 0.361 (0.050) & 5.005 \\
                                                                        & green & 184.478   & 0.367 (0.052) & 2.512 \\
\thickhline
\end{tabular}
\end{center}
\label{results3states}
\end{table}

\begin{table}[ht]
\caption{Results for $J=4$ with corresponding fitted curves in Figures \ref{4statesPLall} and \ref{4statesBall}.}
\centering
\begin{tabular}{c|c|c|c|c}
 \thickhline
Approach & curve  & $\hat{\sigma}^2_j$ & $\hat{p}_j$ (s.e.) & $\lambda_j$ \ \\
 \thickhline
\multirow{4}*{\minitab[c]{Penalized\\ log-likelihood \\ (Fig. \ref{4statesPLall})}} & black  & 50.348 & 0.211 (0.042) & 1.591 \\
 & red & 7.623 & 0.232 (0.050) & 1.213 \\
 & green & 7.245 & 0.244 (0.051) & 0.825 \\
 & blue & 135.968 & 0.313 (0.049) & 0.545 \\
   \hline
\multirow{4}*{\minitab[c]{Bayesian \\  (Fig. \ref{4statesBall})}}
 & black    & 48.376  & 0.223 (0.043) & 3.782 \\
 & red      & 5.473   & 0.221 (0.045) & 4.805 \\
 & green    & 1.106   & 0.198 (0.042) & 5.377 \\
 & blue     & 145.440 & 0.358 (0.048) & 3.355 \\
 \thickhline
\end{tabular}
\label{results4states}
\end{table}

\clearpage


\clearpage


\clearpage

\clearpage

\clearpage

\clearpage

\clearpage

\clearpage

\clearpage

\clearpage


\clearpage

\end{document}